%% file: published_jebo.tex
\documentclass[preprint,12pt,3p,times,review]{elsarticle}
\usepackage{etoolbox}
\makeatletter
\patchcmd{\ps@pprintTitle}{\footnotesize\itshape
       Preprint submitted to \ifx\@journal\@empty Elsevier
       \else\@journal\fi\hfill\today}{\relax}{}{}
\makeatother
\usepackage[utf8]{inputenc}
\usepackage[english]{babel}
\usepackage{amsmath}
\usepackage{amsfonts}
\usepackage{setspace}
\graphicspath{ {figs/} }
\usepackage[ruled,vlined,linesnumbered,resetcount]{algorithm2e}
\usepackage{natbib}
\usepackage{caption}
\usepackage{longtable,multicol}
\usepackage[table]{xcolor}
\setcitestyle{authoryear}
\usepackage{booktabs}

\journal{}

\begin{document}

\begin{frontmatter}

\title{How do governments determine policy priorities? Studying development strategies through spillover networks.}

\author[label5]{Gonzalo Castañeda}
\address[label5]{Centro de Investigaci\'on y Docencia Econ\'omica (CIDE)}
\ead{gonzalo.castaneda@cide.edu}

\author[label5]{Florian Chávez-Ju\'arez}
\ead{florian.chavez@cide.edu}

\author[label1,label2]{Omar A. Guerrero\corref{cor1}}
\address[label1]{The Alan Turing Institute}
\address[label2]{Department of Economics, University College London}

\cortext[cor1]{Corresponding author}
\ead{oguerrero@turing.ac.uk}

\begin{abstract}
Determining policy priorities is a challenging task for any government because there may be, for example, a multiplicity of objectives to be simultaneously attained, a multidimensional policy space to be explored, inefficiencies in the implementation of public policies, interdependencies between policy issues, etc. Altogether, these factors generate a complex landscape that governments need to navigate in order to reach their goals. To address this problem, we develop a framework to model the evolution of development indicators as a political economy game on a network. Our approach accounts for the --recently documented-- network of spillovers between policy issues, as well as the well-known political economy problem arising from budget assignment. This allows us to infer not only policy priorities, but also the effective use of resources in each policy issue. Using development indicators data from more than 100 countries over 11 years, we show that the country-specific context is a central determinant of the effectiveness of policy priorities. In addition, our model explains well-known aggregate facts about the relationship between corruption and development. Finally, this framework provides a new analytic tool to generate bespoke advice on development strategies.
\end{abstract}

\begin{keyword}
Public policy, development indicators, behavioral game, networks, agent-based modeling, Sustainable Development Goals
\end{keyword}

\end{frontmatter}

\section{Introduction}\label{sec:intro}

Throughout the process of economic development, governments prioritize public policies with the aim of reaching specific targets. Such targets may be motivated by internal political agreements, by imitating successful countries or by a broad international consensus (best  practices), among other reasons. Whichever the case, the task of effectively prioritizing policies can be daunting, on one hand, due to inefficiencies (such as corruption) in the implementation process and, on the other, because dealing with a large set of policy goals is not trivial. For example, the recent transition to the Sustainable Development Goals (SDGs) implies that governments should increase their policy spectrum to cover 169 targets as opposed to 18 from the Millennium Project. Moreover, governments have to consider 232 indicators about the relevant policy issues, instead of the 48 previously used \citep{general_assembly_work_2017}.

Leaving aside well-known measurement and data-generation problems, one of the biggest challenges in reaching development goals is accounting for the interdependency between policy issues \citep{nilsson_policy:_2016} (see \cite{pradhan_systematic_2017} for a survey). For example, schooling is likely to exert positive effects on labor markets, so depending on how governments prioritize and coordinate policies, the allocated resources may become complementary or redundant. Furthermore, the policy-issue relationship structure may vary considerably from one country to another. For instance, health policies can have a widespread impact on the socioeconomic indicators of a poor country like Haiti, given the fact that human capital is not a generalized asset in its population. As \cite{nilsson_policy:_2016} put it: \emph{``Implicit in the SDG logic is that the goals depend on each other -- but no one has specified exactly how''}. Some attempts have been made to characterize SDGs as networks of pairwise correlations \citep{le_blanc_towards_2015}. However, it is not obvious how to move from correlations to causal relationships. Even more important, it is not clear how to use such networks for prescriptive purposes without running into well-known limitation of conventional statistical models (see section \ref{sec:literature}).

We propose to think about such interdependencies in terms of positive spillover effects between development indicators.\footnote{The literature on pairwise correlations between SDGs \citep{nilsson_policy:_2016,pradhan_systematic_2017} also considers negative relationships. Here, we focus on positive ones because they have a natural economic interpretation in terms of public policy and in simulating the growth of development indicators. Future extensions might consider negative spillovers as well.} In this network, each node represents a policy issue, and an edge flowing from one node to another symbolizes a spillover from the former to the latter. This network structure can be unique to a particular economy, resulting in a distinctive allocation of resources across the same policy issues that other countries face. Thus, a set of policies that work for a country may be ineffective in another. In addition, there are political-economy considerations that central authorities need to address when allocating resources to different government offices. For instance, in face of imperfect supervision, positive network effects can mask the incompetence of government officials. Even more preoccupying, these situations may elicit incentives to divert public funds for private gains. Together, these mechanisms shape the development strategies observed throughout the world; therefore, building a framework to understand them is paramount. Ideally such a framework could be used to shed some light on the complex process by which countries prioritize public policies and to provide some advice to governments who wish to reach specific targets.

In this paper, we appeal to ideas and tools from behavioral economics and network science in order to develop a new approach to the problem of formulating policy priorities, and to provide a policy-guiding tool. The method builds on a model where a central government assigns resources to different officials who, in the end, decide how much of these resources they will actually use for their original purpose. Three distinctive features define this model: 1) a country-specific network of spillovers (interdependencies) between policy issues; 2) political economy considerations that differentiate policy design from implementation; and 3) a central authority that --through a behavioral game-- achieves development targets by allocating resources while, in parallel, its functionaries learn how much corruption can pass undetected.\footnote{Generally speaking, one can think of inefficiencies in the implementation process. However, in the context of developing countries, the concept of corruption, understood as the diversion of public funds, is highly salient. Therefore, in this paper, we adopt this concept.}

The proposed model allows inferring policy priorities from observed indicators, and to evaluate their suitability for reaching specific targets. With that aim in mind, we consider that, as countries evolve, they leave behind a `development footprint' reflected in their policy indicators. That is, developing countries may use as guides those policies that advanced nations implemented to achieve their current stage of development. In fact, in the study of structural transformations in developing countries, a step-wise development process in which nations follow successful cases is indeed observed \citep{akamatsu_historical_1962}. 

The rest of the paper has the following structure. In section \ref{sec:literature}, we review the literature related to the problem under consideration, and make some comments with regard to the limitations of alternative methodologies. Section \ref{sec:behavioral} introduces the theoretical model and provides a brief computational analysis. In section \ref{sec:data and calibration} we describe the data, its normalization and how the spillover network is estimated. Section \ref{sec:external} presents different tests for the external validation of the model    using information from 117 countries. Then, in section \ref{sec:internal}, we perform internal validation tests, analyzing the outcomes' sensitivity to the model's components (or social mechanisms). Next, section \ref{sec:applications} presents retrospective and prospective analyses for the countries included in the sample. The former allows us to infer the policy priorities that these nations employed throughout the last decade. The latter allows identifying the policy priorities that emerge when following the development footprints of more advanced nations. Finally, we conclude in section \ref{sec:conclusions} with a summary of the empirical results of the model, and provide suggestions for future extensions.

\section{Related literature and alternative methodologies\label{sec:literature}}

The literature studying how a set of policies impacts the economic development of a region (\emph{e.g.}, country, state or municipality) offers different methodological approaches. In this review, we compare three that are frequently considered and one that is closer to ours: econometric analyses, benchmark studies, growth diagnostics and interdependency networks. In particular, we focus our discussion on their main limitations and how our framework can help overcoming them.

\subsection{Regression analyses}

Most econometric-based studies concentrate on linear-regression analysis. A direct implication of considering linear relationships is the implicit assumption of substitutability between public policies. This, in turn, prevents any policy issue from being identified as a binding constraint to the dependent variable (often GDP growth).\footnote{In order to consider parameter heterogeneity, regression-based approaches may employ more sophisticated techniques such as including thresholds in the fitted specifications \citep{minier_institutions_2007}, quantile regressions \citep{canarella_parameter_2004}, estimation of non-linear dependencies through the Generalized Additive Mixed Model, GAMM \citep{wood_generalized_2006}, or pooling countries into different groups in order to limit the degree of substitutability between public policies \citep{lee_both_2009}. These practices, however, are problematic for scaling the number of variables, and give place to numerous specifications. Hence model selection becomes an issue that requires additional methodologies \citep{vuong_likelihood_1989, doornik_statistical_2015}. A similar problem arises when choosing models with interactive terms.}  Hence, the statistical and economic significance of any `independent variable' justifies the use of the associated policy. Another problem arises from estimating average effects. For example, when consultants base their policy advise on these tools, they `forget' that the relationships between the dependent and the independent variables correspond to a hypothetical country with the average characteristics of the data set at hand. Therefore, a country-specific analysis is out of the question. Furthermore, the Rodrik's critique suggests that policies are not random variables, but conscious and strategic decisions made by governments; hence, cross-national data would hardly provide enough variation to assess the relevance of specific public policies in a particular country \citep{rodrik_why_2012}. Finally, it is often the case that policy targets arise from international consensus. Therefore short-term cross-national observations are unlikely independent.

\subsection{Benchmark studies}
Consultants and technocrats tend to articulate their policy advice on issues related to socioeconomic development (\emph{e.g.}, competitiveness, social development, growth, well-being, etc.) by means of benchmark comparisons. For this purpose, they make use of a large set of indicators that describe different realities in the national or international level \citep{rondo-brovetto_comparing_2007,huggins_regional_2010}. By analyzing these indicators in isolation, they establish the minimum standards that laggard regions should attempt to achieve in a subset of selected policies. There are many reasons why this common practice usually leads to erroneous inferences and misleading advice; here we mention a few. First, the absence of a theoretical support does not help to specify how policy priorities should be formulated. Second, there is a high degree of arbitrariness on how the recommended policies should be chosen. Third, this simplistic approach does not take into account the non-linearity and interdependence of public policies. Fourth, this method is unable to evaluate the effectiveness of the recommended public policies.

\subsection{Growth diagnostics}
The growth diagnostics approach aims at identifying the key policy interventions that can ignite growth in a particular region during a specific period \citep{hausmann_growth_2005,hausmann_doing_2008, rodrik_one_2009, rodrik_diagnostics_2010}. It is based on the idea that prices and shadow prices of specific factors (\emph{e.g.}, finance, education, infrastructure and governance institutions) reflect the scarcity of resources. Consequently, it is designed to discover critical bottlenecks in the economy under study. As pointed by \cite{aghion_growth_2009} and \cite{felipe_rethinking_2008}, one of the limitations of growth diagnostics lies in the difficulty of determining a comprehensive list of policy priorities. On one hand, the requirement of expert knowledge in each policy issue makes scalability inviable (\emph{i.e.}, the more indicators the least feasible to implement the analysis). On the other, if distorted prices do not allow identifying binding constraints, a non-price signal has to be assessed (\emph{e.g.}, informal activities). Other limitations include arbitrariness when selecting policy issues and a unidimensional view of policy objectives that focuses solely on GDP growth \citep{Habermann_growth_2011}.

\subsection{Interdependency networks}

The literature on interdependency networks focuses on the interactions between policy issues. The most popular approach uses Bayesian networks to infer how improvements in specific socioeconomic indicators affect policy targets. This is achieved by estimating the network structure of probabilistic dependencies between targets and treated variables \citep{czyzewska_bayesian_2014, ceriani_multidimensional_2016, cinicioglu_exploring_2017}. Besides offering intuitive network visualizations, this methodology allows researchers to infer which policies are the most effective to influence a particular target (\emph{i.e.}, diagnostic analysis), as well as assessing the impact of a particular policy issue on any indicator (\emph{i.e.}, predictive analysis). 

The Bayesian approach described above has, however, several drawbacks. First, similar replicates (\emph{i.e.}, pooled data of ``similar'' countries) are required for the estimation procedure to be feasible because the available time series are usually short for a given country. Thus, when calculating marginal probabilities, this framework is applicable only to sets of countries with an assumed structural similarity. For this reason, it is not possible (at least not with today's available data sets) to analyze countries on a case-by-case basis. Second, this approach makes no attempt to estimate causal relationships and, thus, any interpretation has to be solely expressed in terms of structural dependencies. Third, the data analysis is not backed by economic theory, which makes it a predictive method rather than a policy/prescriptive tool.

Another line of research, based on the idea of building a set of interdependent policies, was developed by \cite{castaneda_complex_2017}, who take a machine learning approach. Here, `policy efforts' are considered the exogenous components of the different development indicators. The authors estimate these efforts through a genetic algorithm that minimizes the distance between previously specified development goals and simulated outcomes (reflected in the level of the associated indicators). The network becomes relevant by allowing spillover effects between policy efforts. The data analysis from this approach, however, lacks a theoretical backbone and suffers from scaling constraints. That is, the larger the set of policy issues, the higher is the dimension of the `chromosome of efforts' that needs to be estimated.

\subsection{Advantages of the proposed methodology}

By means of a behavioral game on a network of policy issues, our method helps to address some of the limitations described above. First, it can handle a very large set of variables (\emph{i.e.}, it is scalable). Second, it accounts for the complex structure of linkages among development indicators (\emph{i.e.}, it does not assume independence between covariates). Third, because it is built on explicit causal/social mechanisms, it is possible to infer how public resources are allocated and diverted (\emph{i.e.}, it allows internal validation). Fourth, it helps us to clarify how the initial conditions, targets and the country's context matter for policymaking (\emph{i.e.}, it produces country-specific estimations). Fifth, it can be used to establish policy guidelines for any particular country when a government establishes a new set of goals (\emph{i.e.}, it is helpful for policy design).\footnote{Although, we do not compute marginal effects for each public issue in this paper, as it is traditionally done in growth regressions, this is possible by running simulations with `deactivated' policies.}

Clearly, one of the biggest advantages of the proposed method is the ability to account for context specificity. The importance of context is particularly salient in the empirical literature, where there is significant evidence showing that countries with similar policy interventions produce very different outcomes \citep{rodrik_one_2009,lee_both_2009}. For instance, improving health and physical infrastructure might be more helpful in lower-income countries than in upper-middle-income ones, while the latter might get more benefit out of public governance and R\&D public policies. This has led many development economists to advocate policymaking based on the identification of the country's binding constraints of growth.

\section{Behavioral game and computational implementation\label{sec:behavioral}}

The output of our model is the simulated evolution of different indicators through the learning process that countries experience as policies are prioritized and implemented. These dynamics are driven by two types of agents: a central authority (or government) and public servants (or functionaries). On one hand, the government allocates resources to different public policies, with the aim of improving the indicators associated to their respective policy issues. On the other, functionaries are in charge of implementing these public policies and, thus, they have incentives to divert public funds for personal gain. Therefore, the incentives of government and bureaucrats are misaligned, giving place to a principal--agent problem.\footnote{Note that we do not assume a benevolent government. For instance, the central authority may try to achieve a set of targets that are in conflict with social welfare.} Furthermore, the interdependency between policy issues encourages free-riding. This is so because positive spillovers mask the real performance of public servants. In this section, we formalize this behavioral game and implement it computationally by means of an agent-based model (ABM).

The behavioral game can be described as a political economy game between the central authority and all public functionaries who have the mandate of implementing policies. Nevertheless, this game has a public-goods flavor since the diversion of funds may have multiplicative adverse effects in societal welfare.\footnote{Our behavioral game does not have the traditional structure of a public goods game, as it considers two types of agents, and initial resources are not assigned exogenously.}  This is due to the fact that smaller spillovers reduce the size of the pie. The intricate interaction between spillovers, diversion of funds, detection of corruption and resource allocation obfuscates the incentive structure of the game; rendering rational modelling inappropriate. Therefore, a more cognitively-viable  specification demands learning-driven behavior where agents adapt to new and limited information via heuristics that are commonly found in laboratory experiments. In order to formalize such heuristics, we employ agent-computing. In the literature of behavioral games, agent-computing has become extremely useful to discriminate among competing explanations of experimental outcomes. This is so because artificial games allow controlling for characteristics that are not easy to isolate in real experiments (\emph{e.g.}, learning, strategic signaling, types of social preferences and interaction networks) \citep{janssen_learning_2006}.\footnote{Three exemplary studies in the application of ABMs are particularly relevant in the context of public goods games:  \citep{bayer_confusion_2009, lucas_effects_2014,reddy_stability_2015}. Examples of ABMs analyzing collective action games can be found in \cite{deadman_modelling_1999, sarin_strategy_2004, castillo_simulation_2005, hichri_emergence_2007, ruttan_economic_2008, chmura_learning_2012, arifovic_behavioral_2011, greiff_learning_2013, nax_directional_2015, ezaki_reinforcement_2016}.}

\subsection{Dynamics of development indicators}

There are $N$ policy issues, each with an indicator that measures its level of development. The level of an indicator depends on how much of the government's allocated resource is effectively utilized in the corresponding public policy. That is, for an amount of resource $P_i \in [0,1]$ allocated to policy issue $i$, the public servant in charge uses $C_i \in [0, P_i]$ effectively in such policy. We call $C_i$ the contribution of the official. Then, $P_i - C_i$ is the amount of public funds that this individual diverts for a personal gain. We refer to this gap as corruption.\footnote{The interpretation of the $P_i - C_i$ gap is, in fact, broader than the idea of corruption. One can think of this gap, for instance, in terms of inefficiencies, since the public servant benefits from shirking and devoting work time to personal activities. For example, an official may prefer to directly adjudicate a government contract to a firm he or she already knows, instead of conducting a proper bidding process --which would imply more work to him or her. In this example, $P_i - C_i$ would represent a loss in efficiency by not hiring the best firm.} In addition to the contribution of the functionary, the level of $i$'s indicator also depends on the public policies of other officials through spillover effects. We model these interdependencies as a network. This network is represented by the adjacency matrix $\mathbb{A}$, where $\mathbb{A}_{ij}>0$ if there are spillovers from $i$ to $j$, and $\mathbb{A}_{ij}=0$ otherwise (the first index denotes rows and the second columns).

Consequently, an indicator is the result of the official's contribution and the spillovers from the contributions of other functionaries. As the government invests in a policy issue, its indicator grows, \emph{i.e.}, the investment accumulates. This means that, if the government has set a target $T_i$ for policy issue $i$, then indicator $I_i$ will reach $T_i$ after $\ell$ periods of investment.\footnote{In this model, a period represents the realization of some events. For example, achieving a target in $\ell$ periods means that the government had to experience $\ell$ events of budget reallocation. A larger $\ell$ implies that reaching the target was more difficult. Therefore, $\ell$ should not be interpreted in terms of time units.} Hence, the dynamics describing the convergence of $I_i$ toward its target is given by

\begin{equation}
        I_{i,t} = I_{i,t-1} + \gamma(T_i-I_{i,t-1})\left(  C_{i,t} + \sum_j C_{j,t} \mathbb{A}_{ji} \right)
    ,\label{eq:propagation}
\end{equation}
where $\mathbb{A}_{ji}$ is the amount of spillovers from $j$ to $i$, $\mathbb{A}_{ii}=0$ and $T_i-I_{i,t-1}$ regulates the velocity of the change in order to reach convergence. Parameter $\gamma$ captures the impact of the effective resources. A simpler version of the model assumes $\gamma=1$. However, calibrating $\gamma$ is useful to exploit the cross-national variation in order to perform aggregate inference.

In this political economy game, the central authority and the public servants solve different problems with limited information. Hence, we model their behaviors mathematically through an adaptive heuristic and directional learning respectively. Then, we describe the computational implementation of the game, and demonstrate its dynamics with an illustrative simulation.

\subsection{Public servants}

We simplify the model by assuming that a government official is in charge of implementing each public policy. Although, we can also think about this agent as an entire office (agency or ministry) that acts through collective behavior. The official's contribution $C_{i,t}$ to the implementation of a public policy depends on how costly it is to divert resources for personal gain. In terms of benefits, the level of the corresponding indicator gives the public servant political status. This, of course, does not depend only on his or her contribution, but also on the spillovers from other policies, (\emph{i.e.}, the contributions of other public servants). The bureaucrat, however, has limited information because he or she cannot directly observe the spillovers (\emph{i.e.}, he or she does not know the network). Instead, the agent evaluates the change in his or her benefits $F_i$. Depending on the evolution of these benefits, the functionary determines its contribution eaach period. First, let us define the benefit function of public servants as

\begin{equation}
    F_{i,t} = (I_{i,t} + P_{i,t} - C_{i,t})(1 - \theta_{i,t}f_{R,t}),\label{eq:benefits}
\end{equation}
where $\theta_{i,t}$ is an indicator function derived from the supervision of the central authority, and $f_{R,t}$ is a function mapping the indicator corresponding to the \emph{rule of law} to a probability. Thus, when the product of these two functions is close to one, the functionary's benefits vanish.

The government cannot measure the real contribution of its public servants, so $P_{i,t} - C_{i,t}$ is not directly observable. However, society generates signals that the central authority might pick up in order to increase supervision efforts in specific policy issues. We assume that the strength of these signals is proportional to the amount of diverted public funds $P_{i,t} - C_{i,t}$. This means that the larger the level of corruption, the more difficult it is to hide, which may cause journalists to uncover them and expose corruption scandals, for example. Therefore, even if the government cannot directly observe the functionaries' contributions, it can target supervision efforts through informed guesses. We model this supervision as a random variable $\theta_{i,t}$. The outcome of this variable is 1 if the public servant in policy issue $i$ is caught diverting public funds, and zero otherwise. Then, the probability mass function of $\theta_{i}$ in period $t$ is

\begin{equation}
    \theta_{i,t} = 
    \begin{cases}
        1 & \text{ with probability } f_{C,t}\frac{(P_{i,t}-C_{i,t})}{\sum_{j=1}^N (P_{j,t}-C_{j,t})}, \\
        0 & \text{ with probability } 1-f_{C,t}\frac{(P_{i,t}-C_{i,t})}{\sum_{j=1}^N (P_{j,t}-C_{j,t})},
    \end{cases}\label{eq:indicator}
\end{equation}
where $f_{C,t}$ maps the indicator corresponding to the \emph{control of corruption} to a probability.

Equation \ref{eq:indicator} implies that officials extracting larger rents are more likely to be caught. Another implication is that small-time corruption, is less likely to be detected. This makes it pervasive and highly unlikely to be eradicated, especially if the country's mechanisms for the control of corruption are weak. 

Note that we introduced two mechanisms through which countries try to mitigate corruption: the rule of law ($f_{R,t}$) and the quality of monitoring efforts ($f_{C,t}$). On one hand, $f_{C,t}$ captures the efforts from the central authority to detect corrupt officials. On the other, $f_{R,t}$ reflects the effectiveness of the state in prosecuting officials who are involved in illicit activities. These two mechanisms describe different constraints that governments face when fighting corruption. For example, in several countries, it is often the case that the central authority improves its methods for monitoring corruption. However, such improvements do not reduce corruption because an ineffective judicial system allows impunity to flourish.\footnote{Note that the model does not assume an `honest' government. In real life, deficient prosecutions are also the result of collusion between the central authority and its functionaries. In these situations, for example, the state attorney's office prepares weak cases, expected to be lost in the courts. This strategy is a proven escape valve to the societal pressures arising from corruption scandals. In the short run, it signals a government that is committed to eradicate corruption; however, in the long run (once the media has lost interest in the case), it reinforces impunity.} This is captured by the interaction $\theta_{i,t}f_{R,t}$, where a corrupt official $i$ might receive a negligible punishment despite being caught diverting funds.

To be more specific, $f_{R,t}$ and $f_{C,t}$ take the form

\begin{equation}
    f_{X,t} = \frac{I_{X,t}}{e^{1-I_{X,t}}},
\end{equation}
where $X=R$ for the rule of law or $X=C$ for control of corruption.

Once we have defined the benefits of the functionaries, we introduce a learning mechanism inspired in \cite{carrella_zero-knowledge_2014}, who applies PID controllers to model firms facing unknown demands.\footnote{PID (proportional–integral–derivative) controllers are realistic behavioral models that aim to capture the learning process of an agent when facing an uncertain environment. Usually, they require three parameters. However, our application does not rely on free parameters.} Then, the public servant updates his or her contribution according to 

\begin{equation}
    C_{i,t} = \min \left\{ P_{i,t},
     \max\left(0, C_{i,t-1} +  d_{i,t} |\Delta F_{i,t}| \frac{C_{i,t-1} + C_{i,t-2}}{2} \right) \right\} \label{eq:contribution}.
\end{equation}

Let us explain equation \ref{eq:contribution} in detail by first focusing on the non-zero element inside the round brackets. The first summand is the contribution from the previous step. The second addend depends on the magnitude of the change in the official's benefits $|\Delta F_{i,t}|$. Factor $d_{i,t}$ is a sign function indicating the direction in which the change of the contribution will go, as suggested by \cite{bayer_confusion_2009}. For this, the public servant evaluates the difference between his or her past contributions $C_{i,t-1}$ and $C_{i,t-2}$, and the difference between his or her past benefits $F_{i,t-1}$ and $F_{i,t-2}$. Incentives to increase the level of the contribution arise from this information. For example, if $C_{i,t-1} > C_{i,t-2}$ and $F_{i,t-1} > F_{i,t-2}$ then the functionary will increase its contribution. Likewise,  $C_{i,t-1} < C_{i,t-2}$ and $F_{i,t-1} < F_{i,t-2}$ incentivize the public servant to increase $C_{i}$. The opposite will occur if any of these inequalities does not hold. More formally

\begin{equation}
    \begin{split}
        \Delta F_{i,t} = F_{i,t-1} - F_{i,t-2}\\
        \Delta C_{i,t} = C_{i,t-1} - C_{i,t-2},
    \end{split}
\end{equation}
from which we define

\begin{equation}
    d_{i,t} = \text{sgn} (\Delta F_{i,t} \cdot \Delta C_{i,t}).
\end{equation}

Going back to equation \ref{eq:contribution}, we have factor $\frac{C_{i,t-1} + C_{i,t-2}}{2}$, which represents the size of the step to be taken when updating the contribution. For consistency, the \emph{min} and \emph{max} functions bound the public servants' contributions.

The official's behavioral component does not require any exogenous parameter, other than the initial conditions. Therefore, it is extremely convenient for empirical applications. Due to the spillovers, this learning mechanism generates co-evolutionary dynamics. Hence, we proceed to model the government, which has to adapt its allocations throughout this co-evolution.

\subsection{Central authority}

The central authority has a vector of targets $\dot{T} = (T_1, \dots, T_N)$ that it wants to achieve for its development indicators. These targets are constant through time. Therefore, the government's problem is deciding how best to allocate its limited resources to different policies in order to reduce the gap between the current indicators and the targets. Formally, the government's problem is 

\begin{equation}
    \min \left[ \sum^N_{i=1} \left( I_{i,t}-T_{i} \right)^2  \right]\label{eq:MSE}
\end{equation}

Equation \ref{eq:propagation} indicates that $I_{i,t}$ is a function of the resource allocation; thus, $P_{1,t}, \dots, P_{N,t}$ are the control variables of the central authority. We call a specific configuration of these variables an \emph{allocation profile}. In addition, the amount of resources that the government can invest is restricted by

\begin{equation}
    \sum_i^N P_{i,t} = B \; \forall \; t.\label{eq:budget}
\end{equation}
Note that a small-enough $B$ guarantees $I_{i,t} \in [0,1]$ and convergence in equation \ref{eq:propagation}. An important characteristic of $B$ is that it reflects the amount of \emph{non-committed} resources of the central authority. That is, a country might have assigned a large fraction of its public expenditure to previously-established purposes such as highway maintenance, agricultural subventions, payment of public debt, etc. Clearly, these expenses are not devoted to transformative policies, so they cannot be accounted for the reordering of policy priorities that the government aims to achieve. In addition, \cite{delavallade_corruption_2006} points out that the majority of diverted public funds come from resources allocated to transformative policies rather than already-committed ones. Thus, empirically speaking, $B$ must be chosen such that it reflects how much budget countries can spare in transforming their economies through public policy.

At each simulation step, the central authority determines an allocation profile and evaluates the gap between the targets and the observed indicators. Due to the budget constraint, whenever the government increases the allocation to one policy issue, it takes away resources from another. Since the government cannot observe the true contribution of its public servants, determining the allocation profile happens through an adaptive heuristic. The amount of resources allocated to policy issue $i$ is determined by 

\begin{equation}
    p_{i,t} = \frac{q_{i,t}}{\sum_j^N q_{j,t}},
\label{eq:allocation}
\end{equation}
where $q_{i,t}$ is the propensity of assigning resources to policy $i$, defined as

\begin{equation}
    q_{i,t} = (T_i-I_{i,t})(K_i+1)(1-\theta_{i,t}f_{R,t}),\label{eq:propensity}
\end{equation}
where $K_i$ is the number of outgoing connections of node $i$, also known as its out-degree.

Equation \ref{eq:propensity} summarizes the intuition of how the government adapts its policy priorities.\footnote{Note that equations \ref{eq:allocation} and \ref{eq:propensity} use a formulation similar to those in reinforcement learning models \citep{dhami_foundations_2016}; however, in this adaptive heuristic, the learning process is informed by corruption dynamics only.} First, the government tries to close the gap $T_i-I_{i,t}$ between the target and the indicator in order to minimize equation \ref{eq:MSE}. Second, $K_i$ is a proxy for how critical a policy issue is. That is, policy issues with a large $K_i$ are central to the development process because they produce spillovers in several other issues; hence, investments in such policies are more effective to reach the indicators' targets. Third, the government tries to reach $\dot{T}$ while, at the same time, attempting to discourage corruption by decreasing its allocation to those policies where the official has been caught diverting funds.\footnote{Empirical evidence suggests that corruption distorts how public spending is allocated. In particular, \cite{delavallade_corruption_2006} shows, with a multinomial model estimated for 64 countries, that corruption reduces expenditure on central pillars of economic development. The argument is that corrupt government officials divert budgetary resources toward sectors where it is easier to get `juicier' bribes and to hide any wrongdoing.} These budgetary changes do not necessarily imply a direct punishment on the public servant. Nonetheless, a budget readjustment is a signal from the government to the bureaucrat asking for discipline and a boost in his or her future contributions. Budget changes are subject to the government capacity to enforce its decisions through the rule of law (that is why $f_{R,t}$ multiplies $\theta_{i,t}$). An alternative interpretation is that a poor rule of law means that central authorities are not truly interested in sending signals when they find indications of corruption.

Finally, the amount of resources allocated to policy $i$ is

\begin{equation}
    P_{i,t} = p_{i,t}B.
\end{equation}

In summary, the model generates endogenous indicators from a political economy game in which policy issues are interdependent. The misalignment between the incentives of the central authority and those of the public servants elicits free-riding and illicit personal gains. To reach its goals, the government penalizes corruption and assigns resources to policy issues with more potential for improving overall economic performance.\footnote{Our methodology considers the problem of corruption from the perspective of the government's budget allocation. We assume that corruption and budgetary decisions are outcomes of a political economy game where public officials have incentives to divert allocated funds. This is consistent with the idea that decentralized forms of corruption are more damaging to the economy \citep{bardhan_corruption_1997}, and with theoretical studies of corruption as a game between non-elected officials and a central government that is accountable to the electorate \citep{accinelli_who_2016}.} The model requires three sources of exogenous information: the targets $T_i$, the network of interdependencies between indicators $\mathbb{A}$ and the budget constraint $B$. All of this information can be obtained from data. 

In the next subsection, we describe the computational implementation of this model and provide a more structured description of the agents' timing. For illustrative purposes, we simulate the dynamics of the three main endogenous variables and interpret their evolution through time.

\subsection{Computational implementation}

The computational implementation of our political economy game consists of instantiating the central authority and each public servant, and letting them determine their allocations and contributions, respectively, through time. Algorithm \ref{algo:yesGovernment} summarizes the agent-based model.

\begin{algorithm}[]
  \KwIn{$\mathbb{A}$, $\dot{T}$, $B$, $\gamma$}
  \ForEach{step $t$}{
       \ForEach{public servant $i$}{
            update contribution $C_{i,t}$;\\
            update benefits $F_{i,t}$;
       }
       \ForEach{node $i$}{
            update indicator $I_{i,t}$;\\
       }
       \ForEach{node $i$}{
            central authority updates $P_{i,t}$;\\
       }
       \If{$|\dot{I}_{i,t} - \dot{I}_{i,t-1}| < \epsilon \quad \forall i$}{
            halt;\\
       }
  }
  \caption{Computational implementation}\label{algo:yesGovernment}
\end{algorithm}

In order to provide an illustration of the model dynamics, we present three outcomes of a typical simulation run on an Erd\H{o}s-R\'enyi random network with 50 nodes and 100 edges. We assume targets $T_i \sim U(0,1)$ and initial resources $P_{i,0}=1/N$ for every $i$. The initial conditions are $I_{i,0} \sim U(0,T_i)$ and $C_{i,0} \sim U(0,P_{i,0})$ for every $i$. 

\begin{figure}[h!]
    \centering
    \includegraphics[scale=.5]{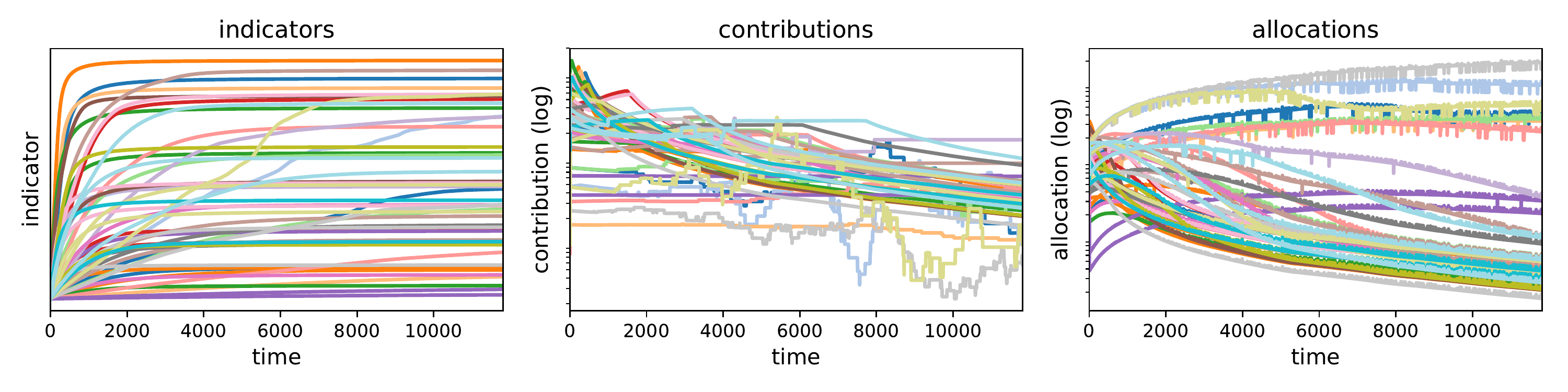}
    \caption{Illustrative model run. This depiction highlights the qualitative nature of the evolution and ordering of different variables across policy issues.}
    \centering\label{fig:evolution}
\end{figure}

Figure \ref{fig:evolution} illustrates, precisely, the evolution of the three main endogenous variables of our ABM. The left panel shows the converging dynamics of the development indicators. Here, we can see that some indicators show a slower convergence speed to their targets. This heterogeneity may be the result of differences between targets and initial conditions, but also from the effects caused by the network topology. The middle panel shows the dynamics of the officials' contributions. It illustrates the process through which each functionary learns the level of corruption where potential penalties from the central authority are bearable. The right panel presents the dynamics of the allocations to each policy issue. The `thickness' of the different lines is caused by the punctuated budget adjustments after corruption has been detected. The crossing of allocations through different steps reflects the process of government adaptation.

\section{Data and network estimation\label{sec:data and calibration}}

In order to perform an empirical application, it is necessary to build a comprehensive database of development indicators across countries. In this section, we introduce a data set of indicators classified into 13 development pillars, the necessary normalizations and transformations, and the method chosen to estimate the network of spillovers between policy issues. Like in any innovative empirical analysis, the quality of the available data and the estimation methods have to improve as knowledge evolves. In particular, revisions of this application are expected since the field of network estimation is rapidly growing. For the time being, we have to make specific methodological choices, as it is the case in the procedure for estimating the direction and weights in the network's links.

\subsection{Data}\label{sec:data}
Our data consists of 79 policy indicators, at the country level, stemming from three different sources. First, we use the data from the Global Competitiveness Report --produced by the World Economic Forum-- which includes a large number of indicators related to economic competitiveness. The second and third sources are provided by the World Bank: the World Development Indicators --for measures on the general socioeconomic development of countries-- and the World Governance Indicators, focusing on topics such as rule of law and government effectiveness among others. The data set consists of annual observations for 117 countries, covering the 2006--2016 period. For 101 countries, we have 11 observations; for 12 countries, we have 10 observations; and four countries have 9 observations (see all countries included in the database in Table \ref{tab:list_countries} from \ref{app:descriptive}).

For a given policy issue $i$, we normalize the respective indicator across countries $c=1, \dots, 117$ and years $y=2006, \dots, 2016$ to an interval between zero (worst possible outcome) and one (best outcome). Formally, this normalization uses the formula

\begin{align}
	I^{\sim c}_{i,y} = \frac{\hat{I}^{c}_{i,y} - \hat{I}_{i,\min}}{\hat{I}_{i,\max} - \hat{I}_{i,\min}}\label{eq:norm1},
\end{align}
where $\hat{I}_{i,\min}$ and $\hat{I}_{i,\max}$ are the smallest and largest empirical values of indicator $i$ across countries and years, and $\hat{I}$ denotes the empirical non-normalized value of the observation. For a small number of highly skewed indicators we use the $4^{th}$ and the $96^{th}$ percentiles rather than the minimum or maximum, respectively.\footnote{If the mean of the computed index using equation \ref{eq:norm1} is below 0.2 (above 0.8), we use the $96^{th}$  ($4^{th}$) percentile instead of the $\max$ ($\min$) function, and then bound the upper (lower) limit to one (zero).} To ensure that, for all indicators, higher values are associated with better outcomes, we apply the inversion $1-I^{\sim c}_{i,y}$ whenever the original index has a negative correlation with the GDP per capita. Finally, missing observations (6.3\%) were imputed using the multiple imputation by chained equations method (see Table \ref{tab:desc_stats} from \ref{app:descriptive} for descriptive statistics on the normalized indicators).

Although we perform our analysis at the level of each country, it is useful to summarize our results by clustering nations by similarity in the structure of their development indicators. In the same spirit, we group indicators into 13 widely accepted development pillars representing broad policy issues that are central for the development of nations. In order to identify clusters of countries, we apply Ward's method, using the L2 (Euclidean) norm as the distance metric across the 79 indicators. The principle behind this method is to group similar countries together by minimizing the variation of the development indicators within the group, while maximizing the differences between groups. We choose this approach over the commonly used income groups defined by the World Bank because it allows us to compare countries in all 79 dimensions rather than just in one.

\begin{figure}[!htb]
	\centering
	\includegraphics[scale=0.7]{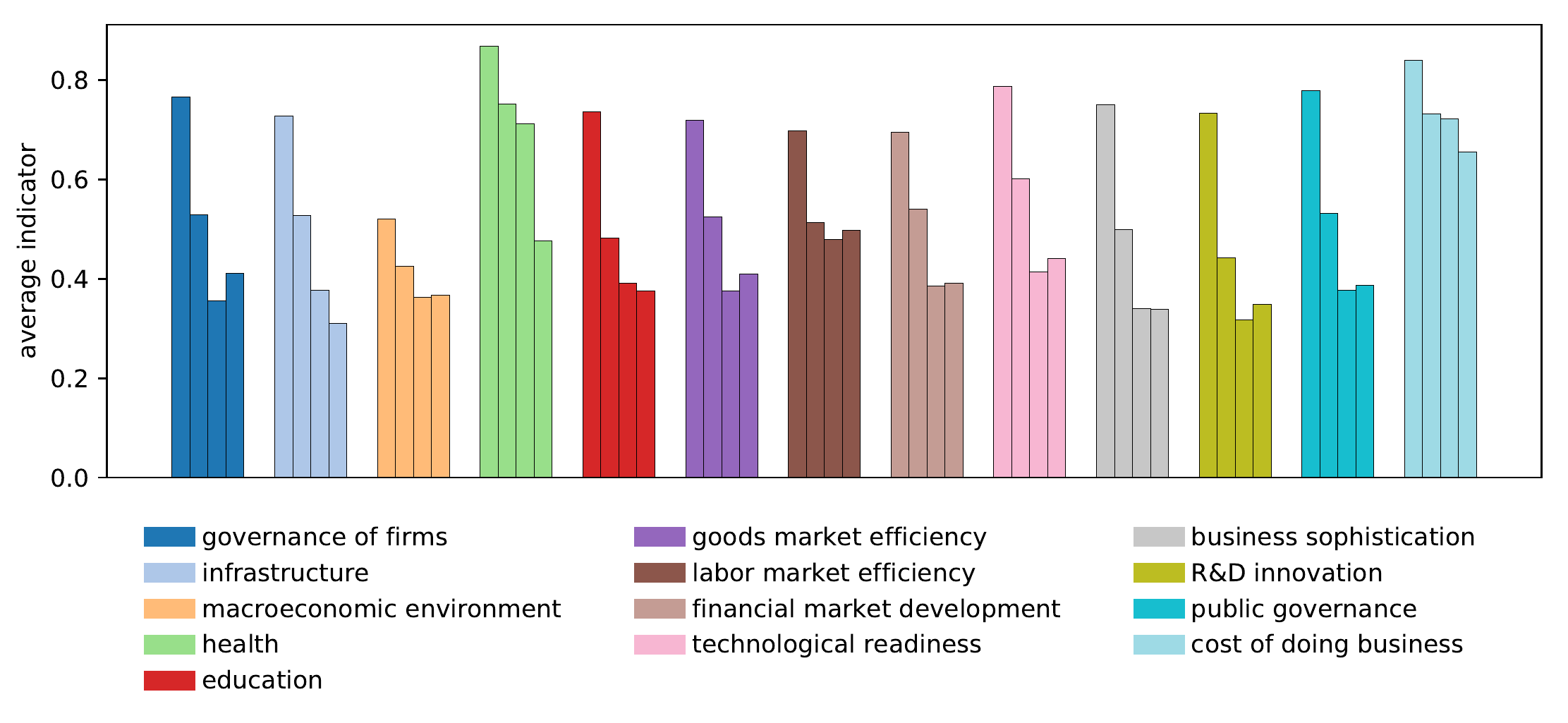}
	\caption{Average levels of indicators by development pillar and cluster. Each colored group represents a development pillar. Within each pillar, the left-most bar corresponds to cluster 1, while the right-most to cluster 4.}
	\label{fig:avg_by_cluster}
\end{figure}

Figure \ref{fig:avg_by_cluster} displays the average level of development in each cluster across the 13 pillars. In general, cluster 1 (the left-most bar within each pillar) contains the most developed countries, while cluster 4 has the least developed ones. Differences between clusters 1 and 2 are sharper in the \emph{education} and \emph{R$\&$D innovation} pillars, while between clusters 3 and 4 differences are more evident in the \emph{health} pillar, but, overall, their discrepancies are less pronounced. In general, the higher the level of a cluster's indicators, the higher the development of the constituent countries. This becomes apparent when looking at the 13 pillars or at the 79 indicators.\footnote{Figure \ref{fig:avg_by_cluster} and Figure \ref{fig:indicators_series} in \ref{app:plots} show the `middle-high development trap'. That is, the gap between clusters 1 and 2 is the largest. Nevertheless, the \emph{health} pillar shows its largest gap between clusters 3 and 4. In addition, indicators such as \emph{infrastructure}, \emph{financial market development}, \emph{technological readiness}, \emph{public governance} and \emph{business sophistication} exhibit significant gaps between clusters 2 and 3.} Accordingly, a successful development strategy should identify the allocation profile that can reach the desired targets in all these indicators.

\subsection{Network estimation}

There exist several methods for estimating directed networks, each with different assumptions and limitations. For example, Bayesian networks elaborated by \cite{pearl_probabilistic_1988,pearl_causal_2016} assume acyclical graphs and do not describe causal relationships,\footnote{Although graphs are used to infer causal relationships by removing and adding edges, the network in itself is not informative about the structure of causalities.} while Granger-causality networks based on \cite{granger_investigating_1969} assume underlying linear relationships between variables as indicated in \cite{castagneto-gissey_dynamic_2014}. Both of these methods require a high observations-to-variables ratio, which is a common limitation in development-indicator data. For all these reasons, we adopt an empirical strategy that has been developed in the estimation of neural networks from functional magnetic resonance imaging data \citep{smith_network_2011,hoyer_causal_2008}. Our estimation strategy is composed of two steps: 1) identifying which pairs of indicators have a significant relationship (and their weights), and 2) inferring the causal direction of these relationships. We apply this strategy to each country.

In order to estimate which pairs of indicators have significant relationships, we apply the method of triangulated maximally filtered graph (TMFG) \citep{massara_network_2017}. This approach is based on the correlation matrix of development indicators. By measuring pairwise correlations --conditional on other indicators (partial correlations)-- a TMFG reveals a meaningful underlying network structure. This network contains information about the complex structure of inter-relationships between policy issues that is not present in the correlation matrix. The TMFG approach is a refinement of the planar maximal filtered graphs method \citep{tumminello_tool_2005}, which was
first developed to identify influential stocks in the US stock market \citep{kenett_dominating_2010}. Once we have obtained the underlying network structure of development indicators, we determine the edges' directions. For this, we follow the method of likelihood ratios developed by \cite{hyvarinen_pairwise_2013}.

The resulting graph is a directed weighted network of development indicators. As we mentioned earlier, network estimation is a very active topic, constantly producing different methods across diverse fields. We have chosen these methods due to their suitability for data sets with few observations and several dimensions, their emphasis on capturing complex structures, and their low computational cost. As the field progresses, we expect more and better alternatives, so our current estimations can be revisited and improved.

\begin{figure}[!htb]
	\centering
	\includegraphics[scale=0.6]{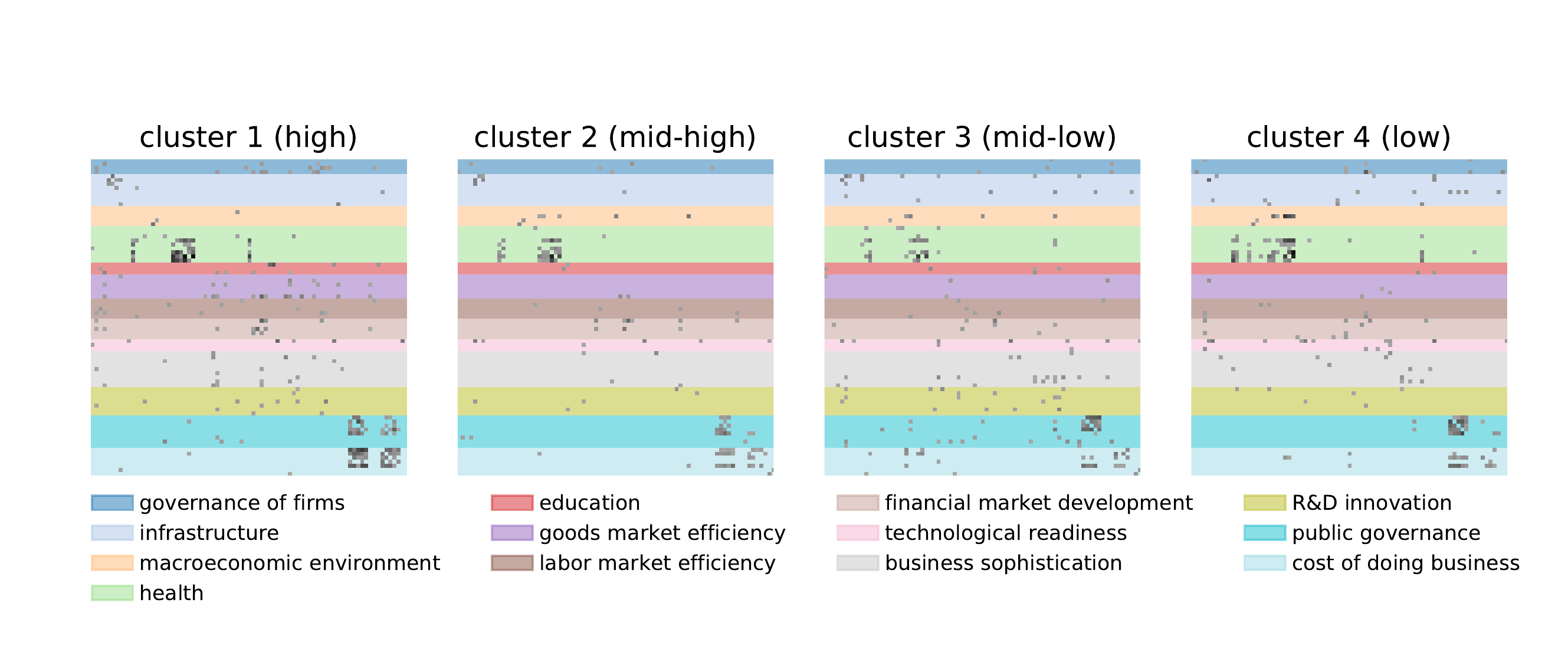}
	\caption{Networks aggregated by cluster. Each panel depicts the aggregated adjacency matrix of each cluster. The nodes (policy issues) have been arranged and colored according to the 13 development pillars. The grayscale dots denote the presence and weight of edges.}
	\label{fig:networks}
\end{figure}

A first approximation to describe topological differences is to show adjacency matrices of aggregated data. For this purpose, we sum the entries of the adjacency matrices across countries in the same cluster. Figure \ref{fig:networks} shows these aggregate networks. Recall that the direction of the spillovers go from rows to columns. The first feature that stands out is that many entries lie near the diagonal. Since we sorted the rows and columns by development pillar, this implies that for some pillars their development indicators tend to have stronger relationships with each other than with indicators from different pillars. However, there is also a substantial amount of non-zero entries outside the diagonal, suggesting that there are policy issues with a significant influence beyond their own pillar. Note that in cluster 1, public governance and cost of doing business are highly interconnected; that, in cluster 2, there is a more sparse network than in the other three clusters; that, in cluster 3, public governance has interdependencies across most pillars; that indicators of education have no connections outside their pillar in cluster 2 and very few in cluster 3 and 4. These facts highlight structural discrepancies in countries that are at different development stages.\footnote{When estimating the networks for each individual country, we also find similarities and differences across their topologies (see Figure \ref{fig:jaccard} in \ref{app:plots}).}

\section{External validation of the model\label{sec:external}}

In the epistemology of ABMs, external validation is usually referred to as the capability of explaining a real-world phenomenon in a satisfactory way. An explanation, in turn, is satisfactory when the phenomenon under study is `grown' by the models' artificial society (\emph{i.e.}, without assumptions about the aggregate behavior of the system). Consequently, a first criterion for external validation is whether our artificial population --composed of government and officials-- is capable of emerging aggregate real-world stylized facts. In this section, we validate the model using empirical estimations corresponding to the sampling period. For clarity of exposition, we provide the details about the estimation procedure in section \ref{sec:estimate_policy}.

\subsection{Cross-national corruption levels}

Our first validation exercise consists in evaluating how well the model can reproduce the cross-national levels of corruption observed in real-world data. The exercise consists of running a sample of Monte Carlo simulations for each country, feeding it with a budget constraint\footnote{The budget constrain is obtained from an indicator of the government expenditure as a fraction of GDP. By using this indicator (as opposed to the absolute expenditure), we take into account the fact that a part of the expenditure is already committed to maintaining the current levels of the indicators.} $B$ and the initial and final values of the indicators in the data set (initial values are the initial conditions and final values are the targets). 

Our empirical measure of corruption comes from an independent development indicator (which was excluded from the estimation): \emph{diversion of public funds}; which is exactly the way in which we define corruption in the model. Our theoretical per-period measure of corruption at the functionary level is the gap $P_i-C_i$. In order to build an aggregate measure for an entire simulation, we sum across its $\ell$ periods. Consequently, we define corruption at the country level through the expression

\begin{equation}
    \bar{D} = \frac{1}{NB}\sum_{i=1}^N \sum^\ell_{t=0} (P_{i,t} - C_{i,t})\label{eq:data_corruption}
\end{equation}
where $N$ is the number of development indicators and $B$ controls for the exogenous cross-national variation originated from the budget constraint.

\begin{figure}[h!]
    \centering 
    \includegraphics[scale=.5]{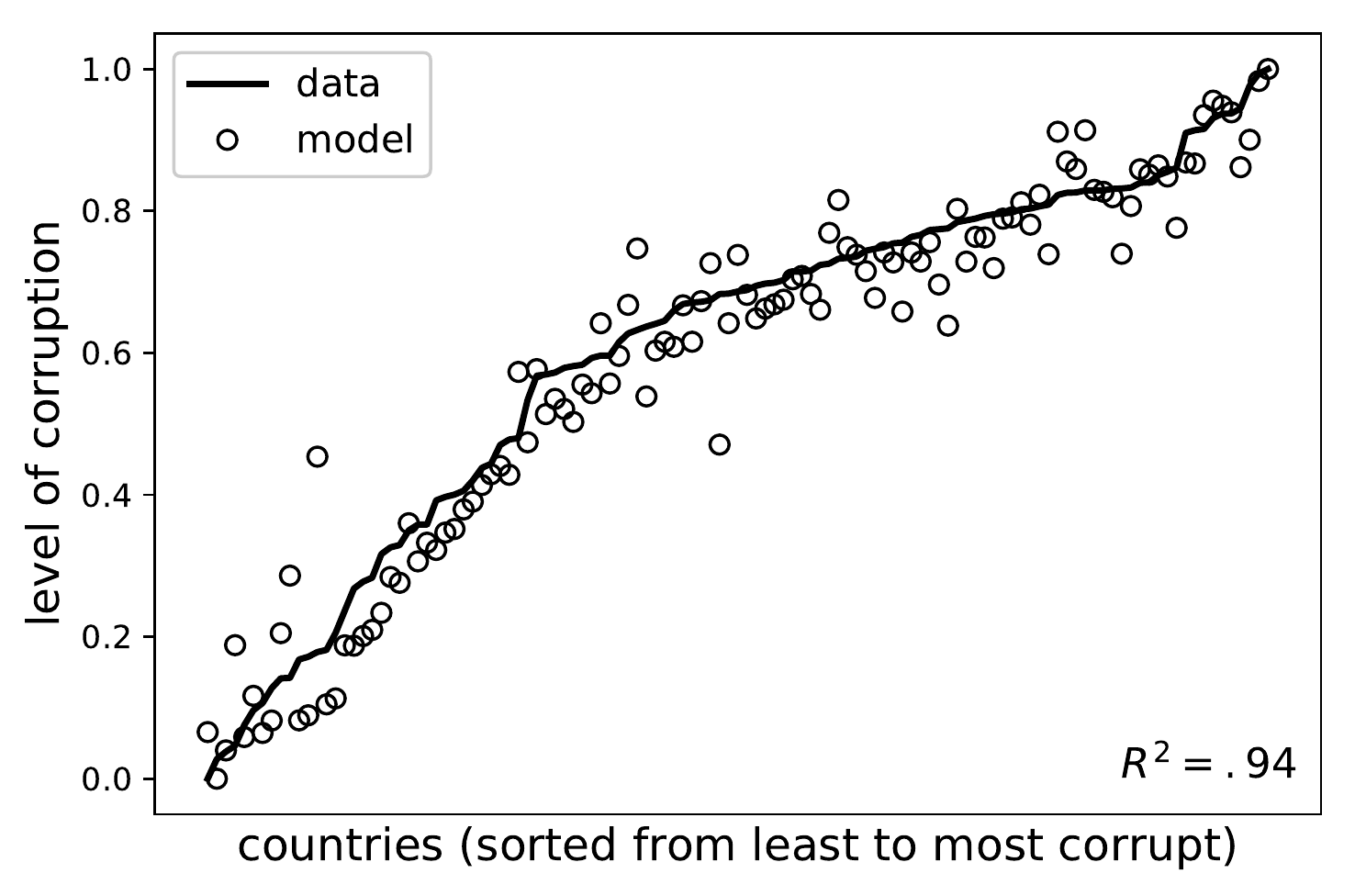}
    \includegraphics[scale=.5]{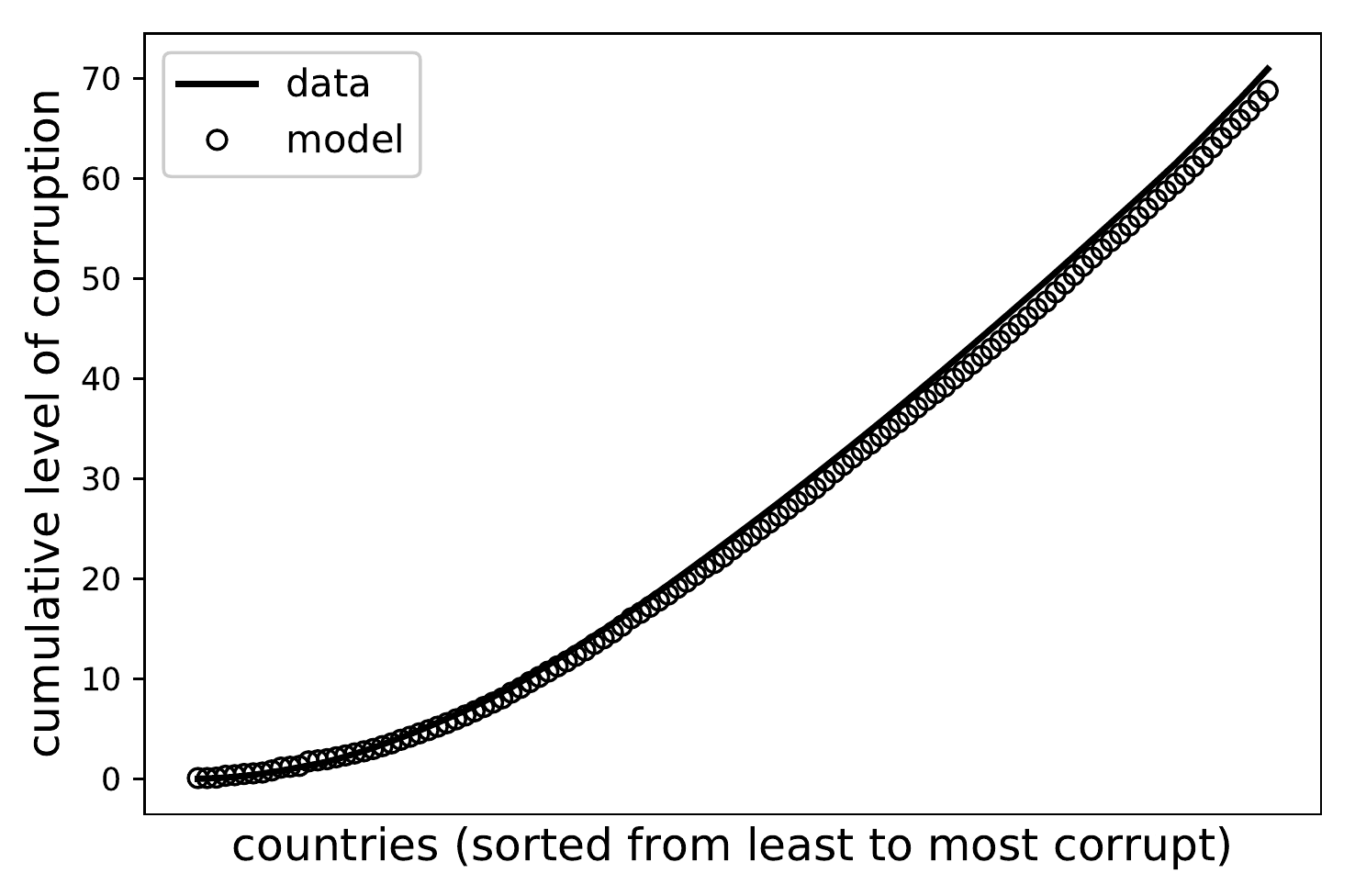}
    \caption{External validation I. Levels of corruption by country. Sampled period: 2006--2016. Model estimation was done by applying a clustering algorithm that classifies countries into different levels of $\gamma$. This procedure seeks to minimize the MSE while controlling for overfitting (see \ref{app:calibration} for details). Each point estimate is the average of 1,000 Monte Carlo simulations. The empirical levels of corruption are measured with the 11-year averages for each country. The $R^2$ is obtained through the classic formula $1 - \frac{\sum(y_i-\hat{y}_i)^2}{\sum(y_i - \bar{y})^2}$.}
    \centering\label{fig:cumulative}
\end{figure}

In summary, we estimate allocation profiles of each country and aggregate the results across indicators to produce comparisons between countries. In order to provide the best possible inference, we calibrate the parameter $\gamma$ across countries (see \ref{app:calibration} for details).\footnote{The qualitative features of our validation also hold for $\gamma=1$}. The left panel in Figure \ref{fig:cumulative} shows the point estimates of corruption for each country and their empirical counterparts. Clearly, the model is able to explain most of the cross-national variation in the observed level of corruption ($R^2=.94$). The right panel shows the empirical and estimated marginal contributions of each country to overall corruption through a cumulative plot. In both cases, Spearman rank-order correlation tests yield coefficients larger than 0.96, providing a first piece of external validation.

\subsection{The corruption-performance relationship}

Extensive evidence shows that the degree of corruption observed across countries is negatively related to their level of economic development (or its performance) \citep{svensson_eight_2005}. Although causality seems to run in both directions, it is commonly argued that economic factors create a demand for better institutions of governance and, thus, for less corruption. Besides this negative relationship, cross-national data shows four additional stylized facts: ($i$) a substantial variation of corruption across countries at the same development stage; ($ii$) such variation is relatively lower among developed countries; ($iii$) a large number of countries with high levels of corruption; and ($iv$) no developing country with a level of corruption similar or lower to that exhibited by the average advance nation.\footnote{To be more precise, we observe practically no overlap between the levels of corruption from laggard countries (clusters 3-4) and from the most developed ones (cluster 1), and a minimal one between clusters 2 and 1.}

In order to empirically measure a country's performance, we calculate the mean of its different development indicators (first across the 11 years of the sample and then across the 79 indicators). For its theoretical counterpart, we average the endogenous indicators from the model, first across the simulation steps and then across indicators as in the expression

\begin{equation}
    \bar{I} = \frac{1}{N}\sum_i \frac{1}{\ell_i} \sum_{t}^{\ell_i} I_{i,t}.\label{eq:data_develop}
\end{equation}

\begin{figure}[h!]
    \centering
    \includegraphics[scale=.5]{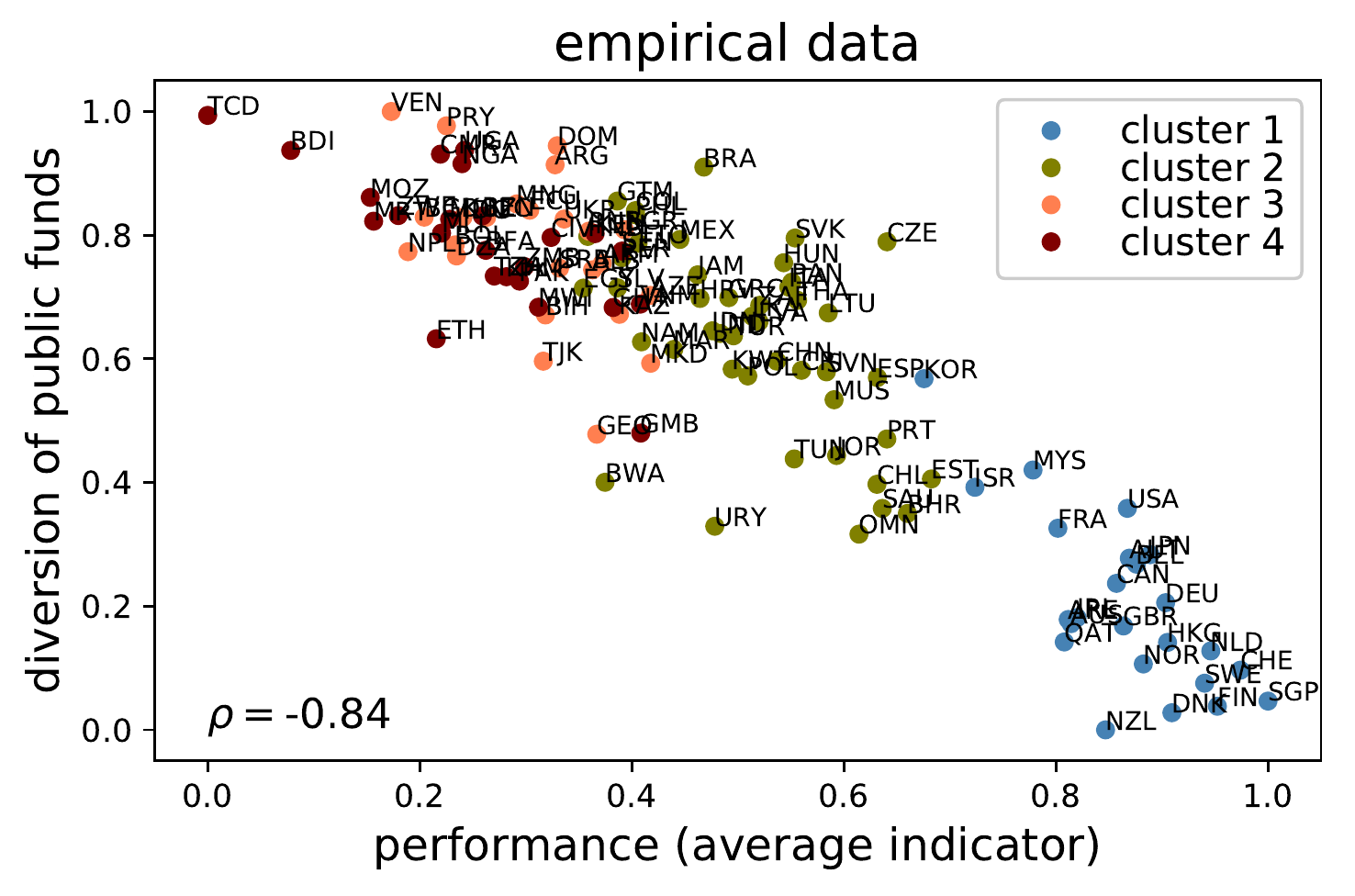}
    \includegraphics[scale=.5]{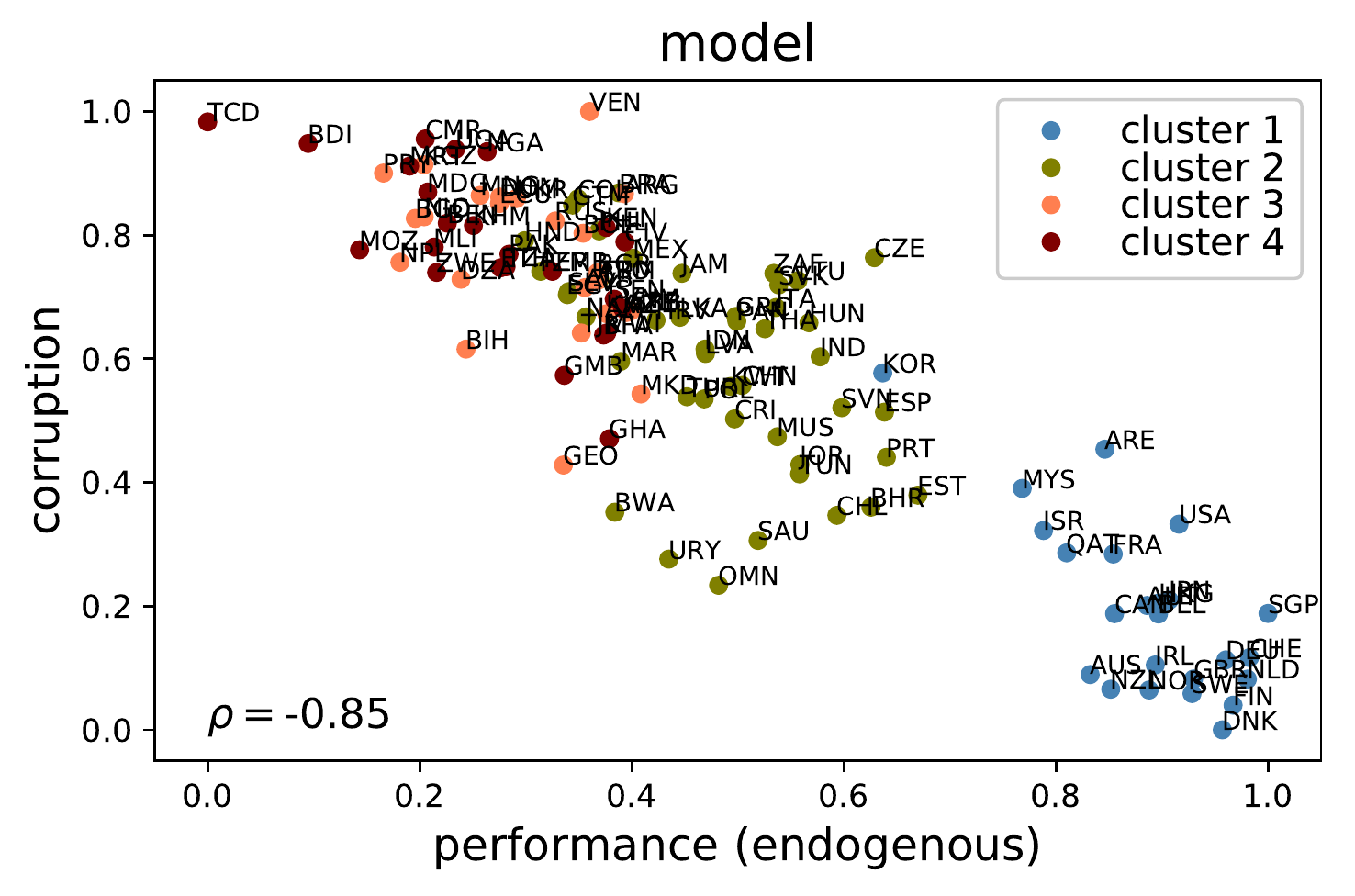}
    \caption{External validation II. Emergence of the corruption-performance relationship. Each dot in the left panel corresponds to an 11-years average for a country. The Y-axis corresponds to the average for diversion of public funds, directly obtained from an indicator in the data set (under the development pillar of \emph{public governance}). The X-axis is the arithmetic mean of the rest of the observed indicators. The coordinates of the dots in the right panel are computed using equations \ref{eq:data_corruption} and \ref{eq:data_develop}. These values were obtained by running 1,000 Monte Carlo simulations for each country. In both panels, each dot has been labeled with its country initials and colored according to its cluster.}
    \centering\label{fig:corruption}
\end{figure}

Note that $\bar{I}$ may be different from the empirical average indicator of a country. This is so because certain indicators converge faster than others, reweighting their inter-temporal averages. Therefore, reorderings in terms of performance are possible, especially since the model is calibrated to minimize the mean squared error (MSE) only with respect to corruption, not performance. The left panel in Figure \ref{fig:corruption} shows the empirical relationship between corruption and performance (the average of all the indicators). The right panel shows the outcome of the model.

The similarity between both panels in Figure \ref{fig:corruption} is remarkable; hence, we can say that the model is informative.\footnote{It must be noted that, although the model is fed with the initial and final values of the empirical indicators, this result is not tautological. This is so because, on one hand, the indicators' values in any period $0<t<n$ are simulated. On the other hand, the model's indicator of corruption is generated through the endogenous variables $C_{i,t}$ and $P_{i,t}$, which are not available in the empirical data set.}  More specifically, the model is capable of emerging the five stylized facts previously mentioned. First, it generates a negative correlation between overall performance and corruption (with a Spearman correlation of -0.85, while the empirical one is estimated at -0.84). Second, countries with similar performance exhibit variation in their levels of corruption. Third, there is heterogeneity in the clusters' corruption variance; for example, cluster 2 (mid-high level of development) has a larger variation than the most developed nations (cluster 1). Fourth, a large amount of countries are clumped in the upper-left quadrant of the plot, which is the area corresponding to low performance and high corruption (all countries in clusters 4 and 3 and some of cluster 2). Fifth, there is practically no overlap between the levels of corruption from countries in cluster 1 and from those in clusters 3 and 4. The ability to emerge all the stylized facts of the corruption-performance relationship validates the relevance of the model's social mechanisms as a whole. To be more precise, these mechanisms are, mainly, the learning process of the public servants, the principal--agent problem related to monitoring and punishment of corruption and the spillover effects through the network of policy issues.

\section{Internal validation of social mechanisms\label{sec:internal}}

The internal validation (or sensitivity analysis) of an ABM has the purpose of detecting whether the model's social mechanisms are relevant for generating its outputs. Such mechanisms are chosen to specify the causal channels that can, arguably, explain the statistical regularities produced by the model. Some mechanisms establish the connection between the agents' decisions and the environment. Other mechanisms specify the interactions between agents and generate societal outcomes. 

In this section, we concentrate on the internal validation of the model's three main components: ($a$) the government's adaptive behavior, ($b$) the bureaucrats' learning processes and ($c$) the spillover network. We validate them by showing their impact on ($i$) the corruption point estimates, ($ii$) the corruption-performance relationship, ($iii$) the public servants' incentives to contribute, and the ($iv$) configuration of the top 10 priorities in allocation profiles.

Internal validation is achieved via sensitivity analysis, which consists in studying the model's outcomes under different specifications. In each specification, we `turn off' one of the components. To turn off government behaviour, we replace the adaptive heuristic by a random choice $P_i \sim U(0,1)$ (normalizing to make sure they add up to $B$). For the public servants, we replace the heuristic for learning $C_i$ with a uniformly-distributed random choice in $[0, P_i]$. Finally, to deactivate the spillover effects, we replace $\mathbb{A}$ with a weighted identity matrix in equation \ref{eq:propagation}, while leaving everything else intact. Let us refer to the original specification as the \emph{full model}. When a simulation outcome differs from the one of the full model, we say that the component or mechanism is relevant to explain it. Something distinctive about internal ABM validation is the ability to perform tests at the both micro- and macro-levels. This is so because, here, every behavior and interaction is explicit, and their outcomes do not rely on assumptions about coordination or aggregation. Hence, the evidence that we present in this section goes beyond the usual standards of alternative empirical approaches.\footnote{The results from the sensitivity analysis also hold for the parameter-free version of the model ($\gamma=1$), reinforcing the validity of the chosen social mechanisms.}

\subsection{Sensitivity of cross-national corruption levels}

\begin{figure}[h!]
    \centering
    \includegraphics[scale=.5]{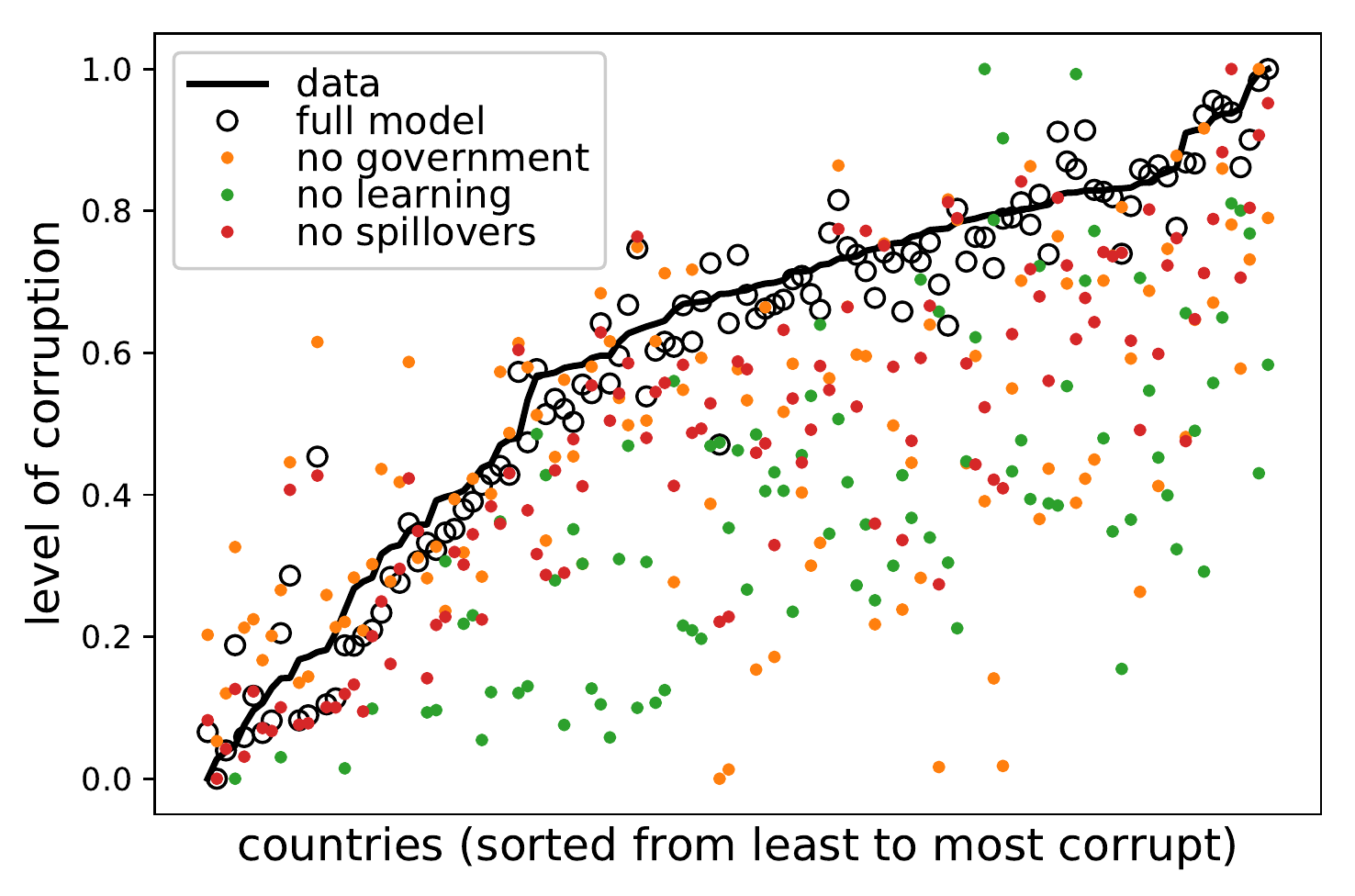}
    \includegraphics[scale=.5]{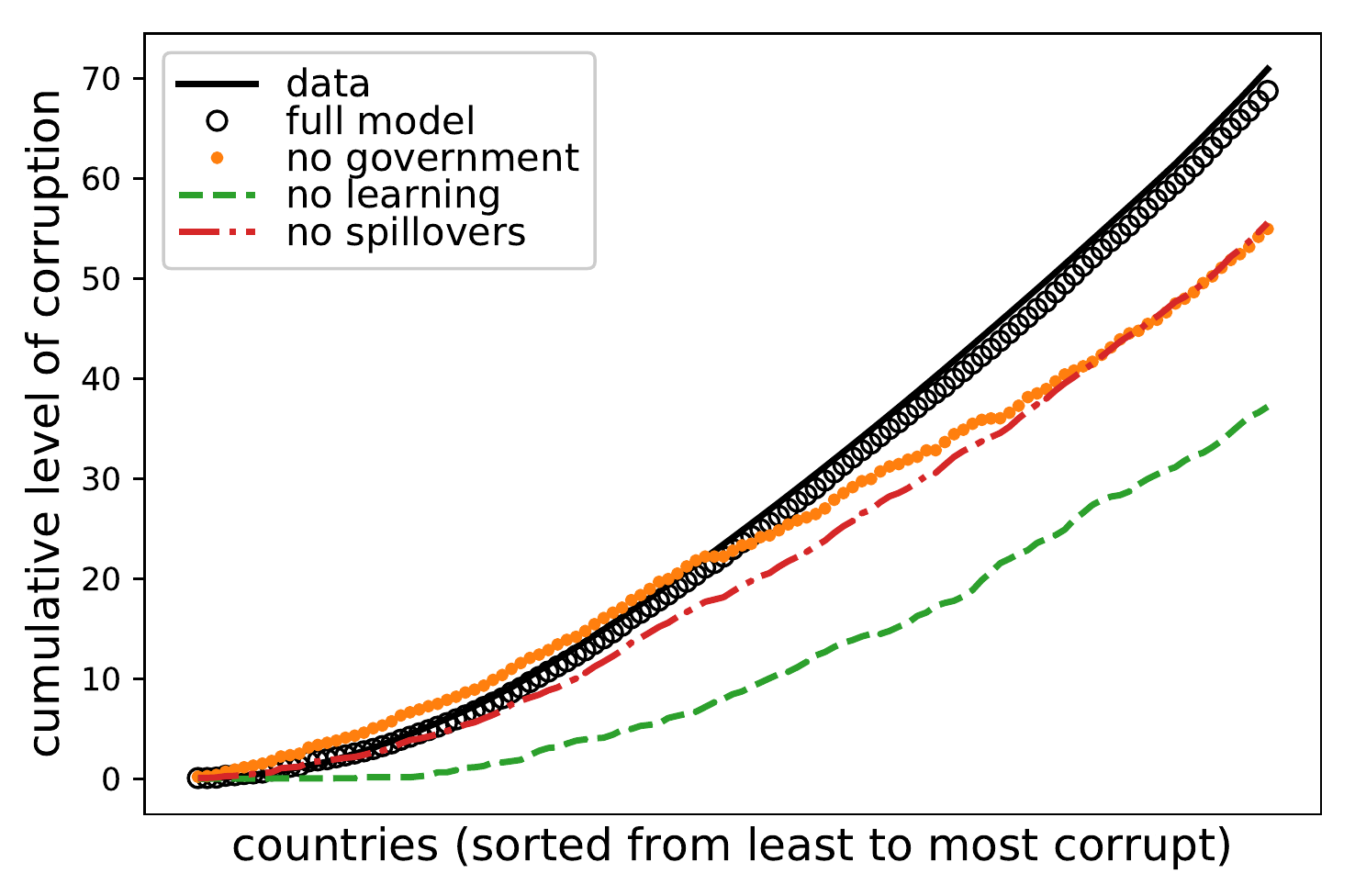}
    \caption{Internal validation I. Sensitivity of cross-national corruption levels. Left panel: point estimates. Right panel: cumulative level of corruption.}
    \centering\label{fig:sensi3}
\end{figure}

Figure \ref{fig:sensi3} shows the estimated levels of corruption without each of the main model components. The left panel shows the point estimates. Confidence intervals have been computed for each point estimate of the full model (too narrow to show them visually), suggesting that, in most cases, deactivating the respective mechanism produces a significant deviation from the estimate. The right panel shows the same data in its cumulative form. Clearly, each of the three components has an important effect in the countries' marginal contributions to overall corruption.

\subsection{Sensitivity of the corruption-performance relationship}

\begin{figure}[h!]
    \centering
    \includegraphics[scale=.45]{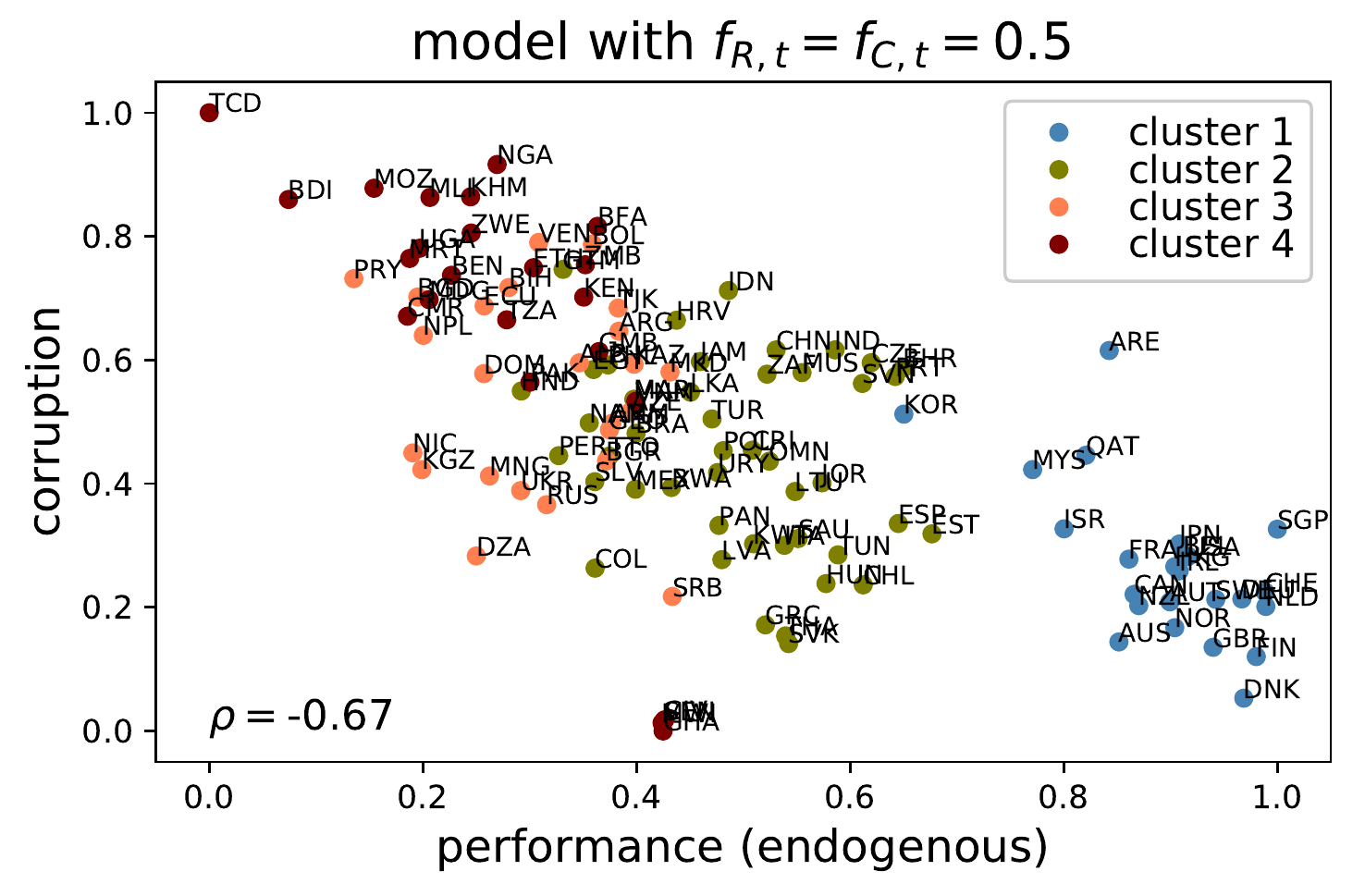}
    \includegraphics[scale=.45]{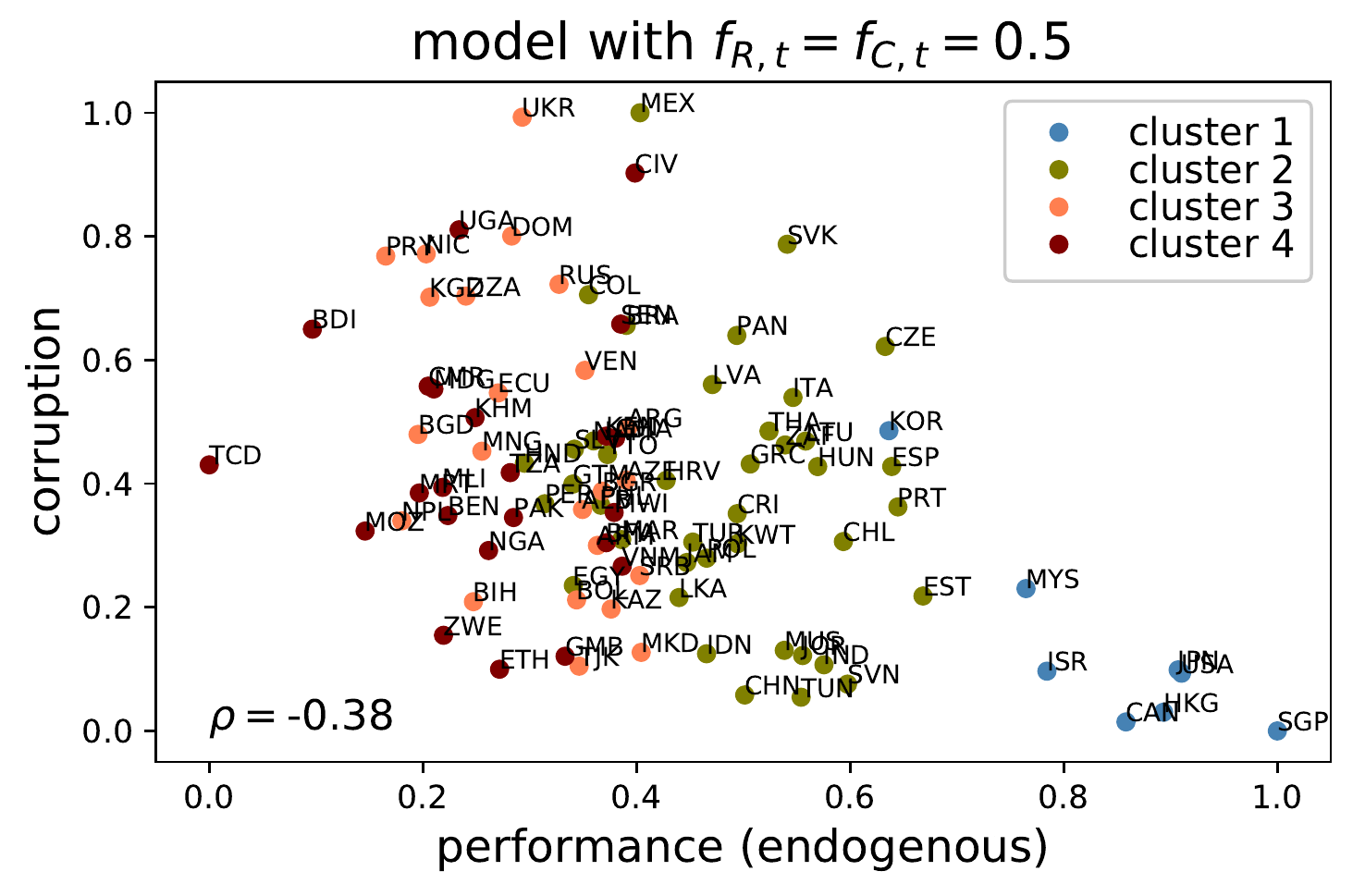}
    \caption{Internal validation II. Sensitivity of the corruption-performance relationship. Right panel: random functionaries' contributions. Left panel: random government's allocations.}
    \centering\label{fig:sensi4}
\end{figure}

Figure \ref{fig:sensi4} shows a significant distortion of the corruption-performance relationship when deactivating the government or the public servants. In both cases, the correlation decreases significantly. In contrast, there is no apparent change when the spillovers are removed (see Figure \ref{fig:sensi3b} in \ref{app:corrNoNet}). Given that the network provides a way to deal with the interdependence between policy issues, this last result is intriguing and deserves further analysis, which we elaborate in section \ref{sec:estimate_policy}.

The model seems particularly sensitive to the learning mechanism of the public servants. We know, from the equations of our behavioral game, that the variables of \emph{rule of law} and \emph{quality of monitoring efforts}, are important for the agents' learning process. Hence, both components influence the emergence of the stylized facts in this relationship. Nevertheless, the inclusion of these variables is not sufficient to produce the observed corruption-performance pattern. For example, we perform simulations holding monitoring and punishment efforts fixed and find that, for $f_{R,t}=f_{C,t}=.5$, the relationship is still negative but the model overestimates corruption in low-corruption countries and underestimates it in high-corruption ones (see \ref{app:corrNoNet} for details).

\subsection{Sensitivity to the spillover network}\label{sec:sensi_network}

So far, we have shown that the network affects the estimates on corruption, but not the corruption-performance relationship. While these tests only consider aggregate stylized facts, there exists a variety of tests that can be performed at the micro-level. Put differently, we have only explored the role of the spillovers at a cross-national level, while its most relevant effects occur within countries. For instance, different nodes are expected to exhibit different outcomes depending on their connectivity. Perhaps the clearest way to show this is by following our argument on how ``\emph{positive network effects can mask the incompetence of the government officials}'' (see section \ref{sec:intro}). In terms of the model's outputs, we would expect that, \emph{ceteris paribus}, nodes with more incoming spillovers would contribute less.

\begin{figure}[h!]
    \centering 
    \includegraphics[scale=.5]{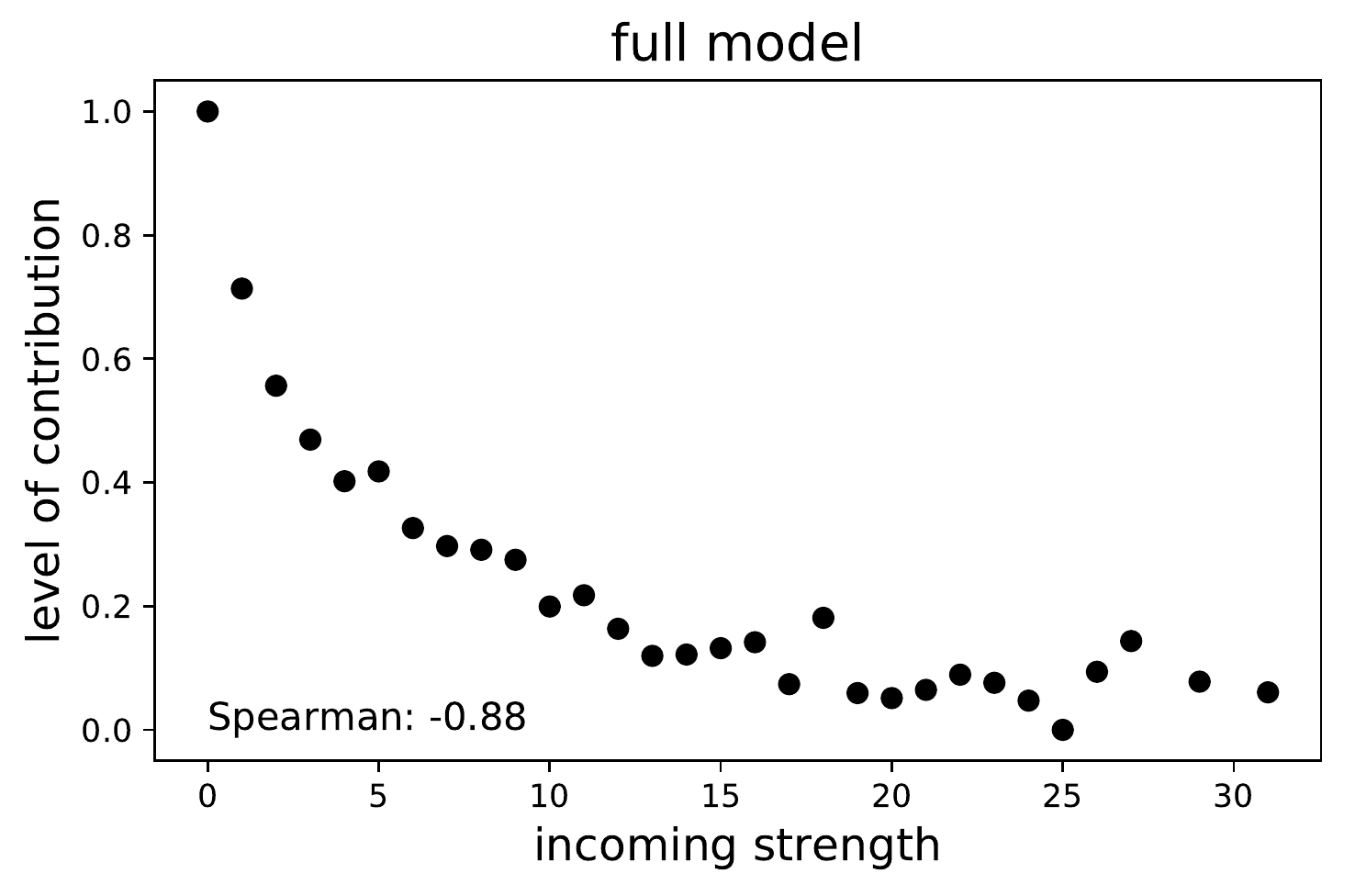}
    \includegraphics[scale=.5]{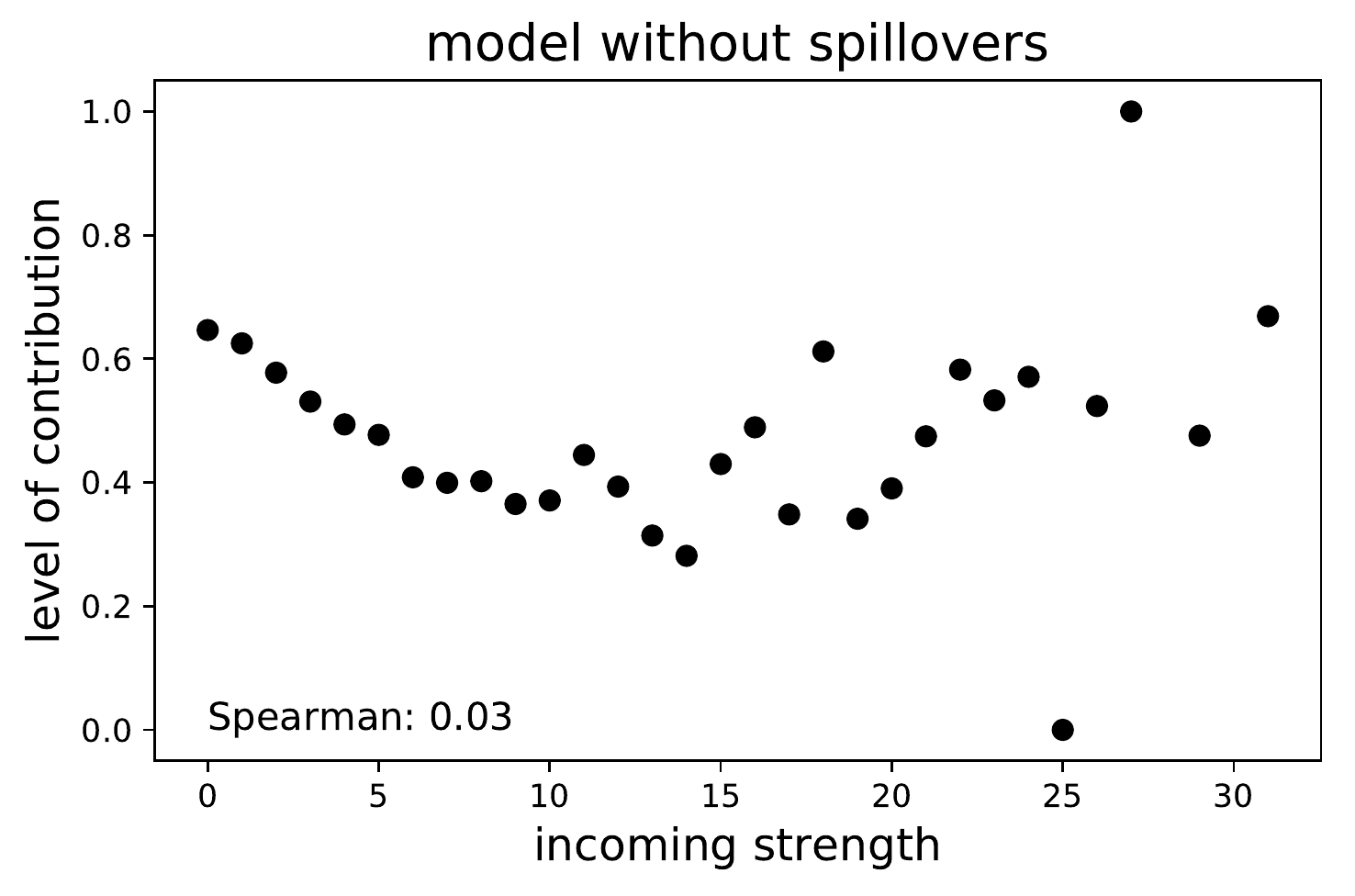}
    \caption{Internal validation III. Relationship between incoming spillovers and contributions. Left panel: complete model. Right panel: model without spillovers. The dots correspond to contributions averaged across nodes that fall in the same bin of incoming strength. The unit of analysis is country-node and the value of the observation is the inter-temporal average.}
    \centering\label{fig:outStrength}
\end{figure}

The left panel in Figure \ref{fig:outStrength} shows the relationship between incoming spillovers ($\gamma\sum_j\mathbb{A}_{ji}$) and contributions ($\frac{1}{\ell_i} \sum_{t}^{\ell_i} C_{i,t}$)  at the level of each node $i$, calculated from Monte Carlo simulations over the entire data set. Different contributions were averaged across nodes with a similar amount of incoming spillovers (\emph{i.e.}, they were binned) --also known as incoming strength. In order to demonstrate the effect of the network, the right panel shows the same output, but for the model without spillovers. Clearly, removing the network mitigates the negative relationship between incoming spillovers and contributions. In fact, statistically speaking, the relationship for the model without a network is negligible (the Spearman correlation yields a p-value of 0.85). Accordingly, these simulation outcomes are consistent with the theoretical foundations of our model.

\subsection{Sensitivity of the government's top priorities}\label{sec:priorities}

\begin{figure}[h!]
    \centering
    \includegraphics[scale=.35]{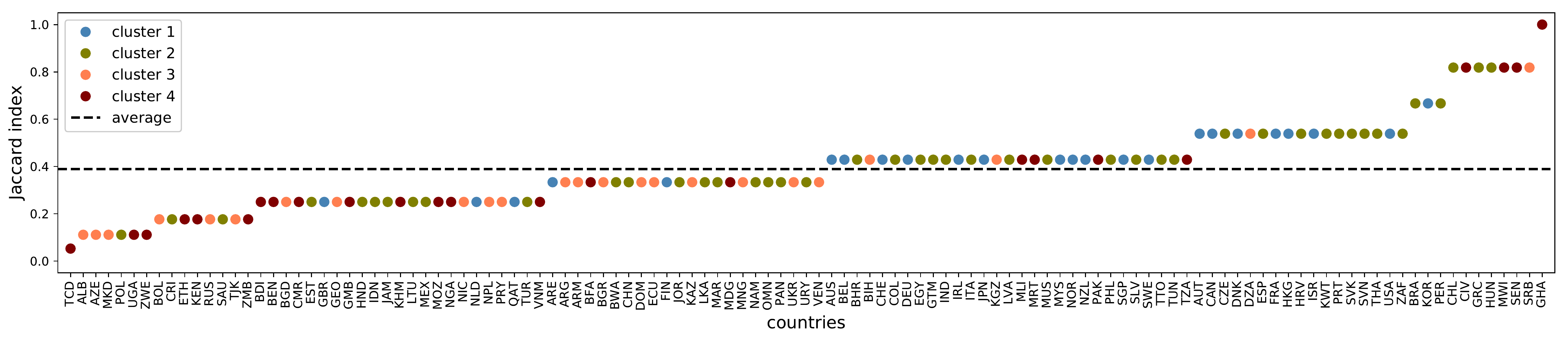}
    \caption{Internal validation IV. Similarity of the top-10 priorities (Jaccard index). The top priorities are the policy issues that received the most resources from the allocation profile.}
    \centering\label{fig:sensi4b}
\end{figure}

Perhaps the most important effect of the network is in the estimation of the allocation profile. That is, when a consultant ignores the spillover effects, he or she may recommend erroneous policy priorities. To illustrate this, let us consider the top 10 policy issues in the estimated allocation profiles. If the network affects this estimation, removing it should yield a significantly different set of top-10 policy issues. We evaluate this difference through the Jaccard membership index. Here, a value of 1 means that the top 10 priorities are the same with or without network (regardless of the order), while 0 means that they are entirely different. Figure \ref{fig:sensi4b} shows a systematic discrepancy between the top-10 priorities with and without spillovers. In particular, we obtain an average Jacccard close to 0.4 across all countries; thus, the highest allocations differ considerably when we discard the spillover effects from the analysis. These results speak not only of the relevance of the network, but of the importance of considering country-specific contexts. Through other methodologies, this is not obvious because one has to work with aggregated data and average effects. Therefore, this exercise demonstrates the strengths of a computational approach in a problem where country-specificity is important for policy prescriptions.

\section{Applications}\label{sec:applications}

In this section we present, firstly, the general results of a retrospective analysis where we infer the average allocation profiles that countries used during the observation period. Secondly, in a prospective analysis, we present two types of Monte Carlo simulations to illustrate the applicability of our framework for guiding policy-making. On one hand, we infer the policy priorities of three country cases, as if they would imitate more advanced nations. Then, on the other, we infer the most feasible development mode for each developing country in the sample ($\emph{i.e.}$ those outside cluster 1).

\subsection{Estimating past policy priorities}\label{sec:estimate_policy}

In order to infer the allocation profiles used in the past decade, we assume that the targets for the set of development indicators coincide with the values observed at the end of the sampling period. Let us look into the estimation of allocation profiles. $I^{\sim}_{i,t}$ denotes the empirical development indicator of policy issue $i$ during the $t^{th}$ period of the data set. For a given country, the exercise consists of running the model using the estimated network, initial conditions $I_{i,0}=I^{\sim}_{i,0}$, targets $T_i=I^{\sim}_{i,n}$ (where $n$ is the last observation in the data), and a budget constraint $B$. Each simulation $m$ generates $N$ time series of the form $P^m_{i,0}, \dots P^m_{i,\ell_m}$ where $\ell_m$ is the end period of simulation $m$. We obtain the average allocation $\bar{P}^m_{i} = \frac{1}{\ell_m}\sum_{t=0}^{\ell{_m}} P^m_{i,t}$ in simulation $m$ by computing the inter-temporal average of its time series. Finally, we compute the mean $\bar{P}_{i} = \frac{1}{m}\sum_{m=1}^M \bar{P}^m_{i}$ across simulations. For presentation purposes, we group this information into clusters and development pillars, and compute averages.\footnote{The initial values of $F_{i,0}$, $F_{i,-1}$, $C_{i,0}$, $C_{i,0}$ and $P_{i,0}$ are randomly determined in each simulation. However, sensitivity tests show that the estimated allocation profiles do not change significantly with the initial conditions of these variables.}

\begin{figure}[h!]
    \centering
    \includegraphics[scale=.7]{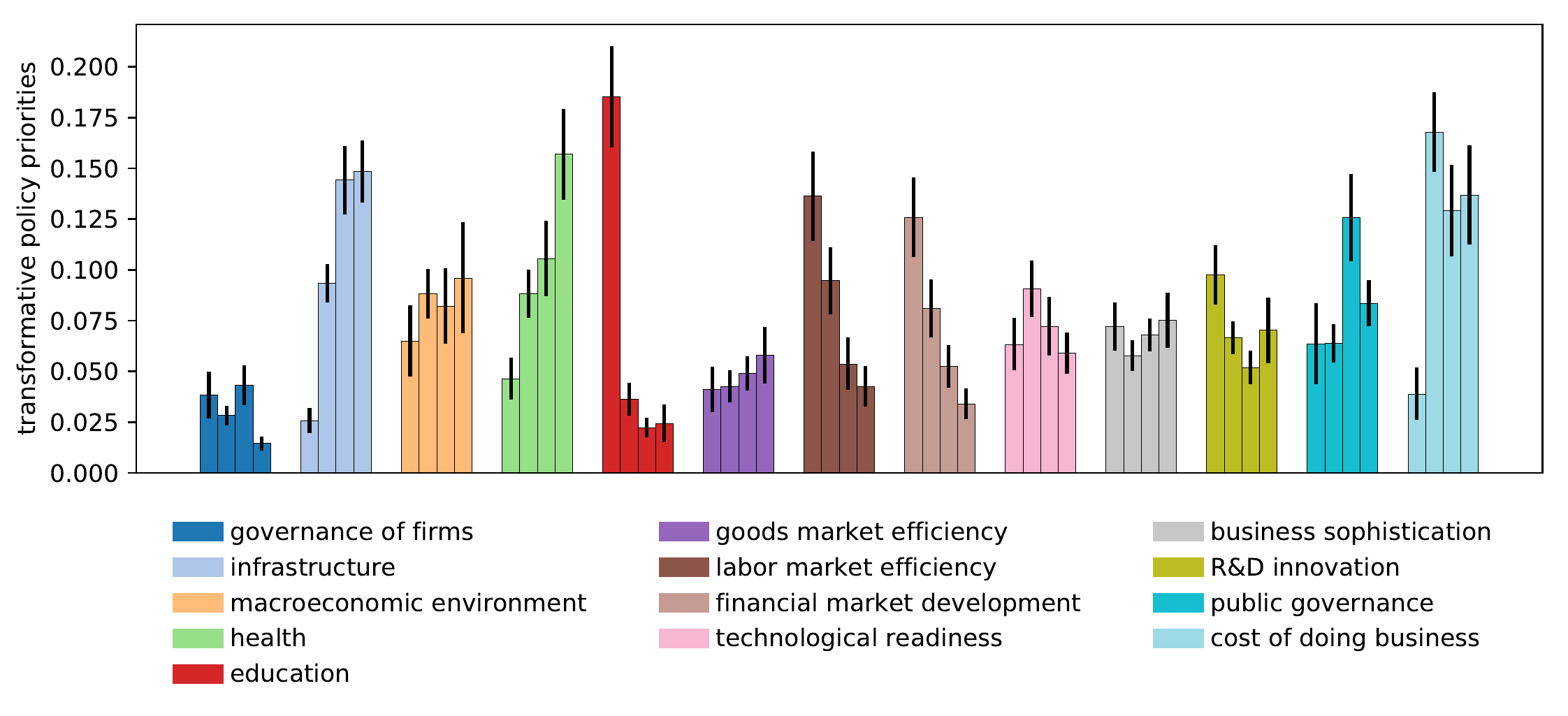}
    \caption{Average allocation profiles by cluster and development pillar. Each color represents a development pillar. Within each pillar, the left-most bar corresponds to cluster 1 and the right-most to cluster 4. The vertical lines on top of the bars denote the standard errors, computed from the cross-national variation. Number of simulations: 1,000 per country.}
    \centering\label{fig:results1}
\end{figure}

Figure \ref{fig:results1} presents the average allocation profiles at the level of development pillars and clusters. Six important results emerge from this result:

\begin{enumerate}
    \item Within each cluster, policy priorities are not uniform across development pillars.
    \item Each cluster has a different ordering of their policy priorities.
    \item Policy priorities are not dictated by the average indicators (Figure \ref{fig:avg_by_cluster}), which means that the model captures the transformative character of the allocation profiles.
    \item On average, countries in clusters 3 and  4 prioritized \emph{infrastructure}; in cluster 2, \emph{cost of doing business}; in cluster 1, \emph{education}.
    \item On average, the lowest priority for countries in cluster 4 and 2 was given to the \emph{governance of firms}; in clusters 3 to \emph{education} and in cluster 1 to \emph{infrastructure}.\footnote{Recall that we are only measuring transformative changes, not total budget allocations. Hence, while a country like Germany may have the highest committed expenditure in infrastructure, it does does not need to spend a substantial amount in transforming it.} 
    \item Transformative policies across clusters were very different in the following pillars: \emph{labor market efficiency}, \emph{health} and \emph{financial market development}. On the contrary, they were relatively similar in the following pillars:  \emph{business sophistication}, \emph{goods markets efficiency} and \emph{technological readiness}.  
\end{enumerate}

In brief, it can be argued that, during our sampling period, the empirical evidence shows that countries followed different paths depending on their development stage. Likewise, the estimations of policy priorities indicate that countries did not set them up in terms of the development gaps observed between pillars; for example, \emph{public governance}, \emph{R\&D innovation}, \emph{business sophistication} and \emph{education} in cluster 2 were not prioritized despite their large gaps with respect to cluster 1. Altogether, the model is capable of capturing the wide spectrum of policy priorities for the countries in the data set and, hence, it has the potential to discover new transformative allocation profiles when different targets are established.

\subsection{Following development footprints}

In 1962, the Japanese economist Kaname Akamatsu conceived the theory of the `flying geese' to describe a `catch-up' development process, observed in Asian countries during the 20th century \citep{akamatsu_historical_1962}. His theory was inspired by the observation that Asian economies developed according to an inverted-V pattern, like wild flying geese. In other words, advanced economies move to more sophisticated industries while developing countries become competitive in those industries left behind.\footnote{Recent evidence of this process and the creation of comparative advantages in Asian countries can be found in \cite{ozawa_japan-born_2011}.}

The flying geese phenomenon is pervasive among countries that have developed successfully since the industrial revolution \citep{lin_flying_2013}. According to \cite{lin_comparative_2013}, the catch-up process was possible because these countries targeted mature industries from advanced countries with similar factor endowments and a relatively close GDP per capita. Hence, a latent comparative advantage becomes manifest when a country undertakes important changes in their physical and institutional infrastructure. Such changes occur when policies are implemented to deal with binding constraints and when the governance architecture is modified to handle information and coordination failures. In other words, the `flying geese' of industrial transformation becomes viable because countries moving up in the development ladder have target nations in mind when implementing transformative policies. For this reason, policy indicators can be seen as development footprints that guide countries when climbing up the ladder. We say that a country adopts a `development mode' when it follows the footprints of another country (\emph{e.g.}, Argentina adopts the French development mode when the former targets the indicators of the latter).

If, indeed, a country decides to adopt a `development mode', the question is: how should the central authority prioritize public policies? Our framework provides a method to address this question. Consider country $x$ and its vector $\dot{I}^{\sim x}_{n}$ of observed development indicators. Suppose $x$ wants to follow the path of country $y$. Then, $x$'s targets become $\dot{T}^x=\dot{I}_n^{\sim y}$. The average allocation profile resulting from applying the model to this data is informative about the policy priorities that $x$ would pursue if it was to adopt $y$ as its development mode.

According to empirical evidence, emerging economies tend to follow modes that are, somehow, similar in terms of their productive and social capabilities. In the context of our analysis, we can think of this notion of proximity as the four clusters. Hence, we assume that members of cluster 4 try to follow countries from cluster 3, and these ones chase those in cluster 2, and so forth. Let us illustrate this by looking at three countries: Mexico in cluster 2, Albania in cluster 3 and Nigeria in cluster 4; and assuming that they try to follow the development footprints of two countries from the cluster above theirs.

\begin{figure}[h!]
    \centering
    \includegraphics[scale=.65]{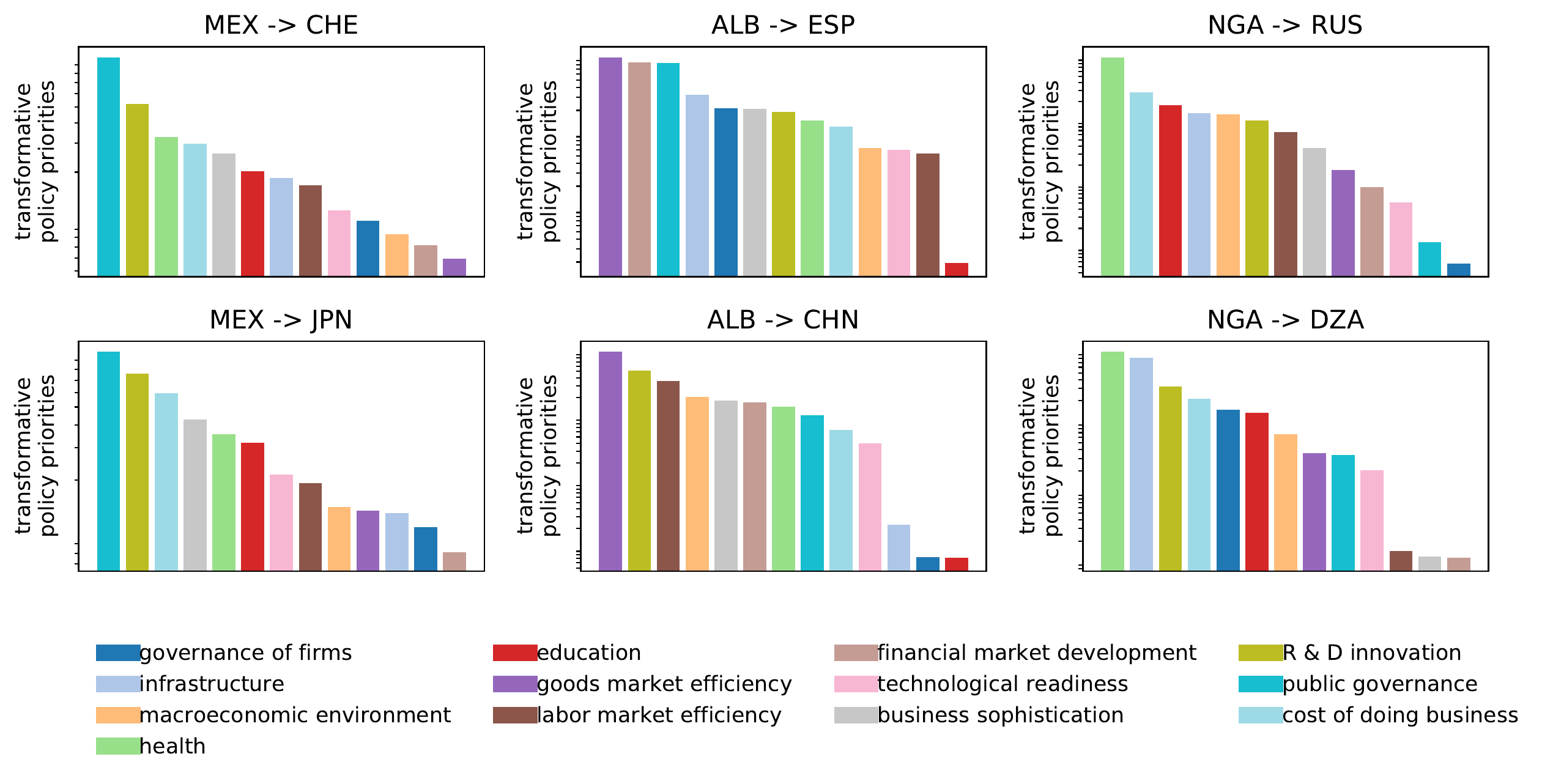}
    \caption{Average allocation profiles of three country cases. Number of simulations: 1,000 per pair of countries.}
    \centering\label{fig:cases}
\end{figure}

Figure \ref{fig:cases} shows the average allocation profile of each catch-up process. The panels present the results for each of the three countries when trying to follow two more-advanced countries (\emph{i.e.}, development modes). In the left panels, we can see that the pillars of \emph{public governance} and \emph{R\&D} are the two top  priorities for Mexico, regardless if it tries to follow Switzerland or Japan. However, depending on the development footprints that Mexico tries to follow, other policy issues such as \emph{infrastructure} occupy a very different position in its allocation profile. This reordering of priorities becomes more obvious in the case of Albania. For example, if Albania attempts to follow Spain, its second priority should be the development of financial markets. However, this policy issue becomes Albania's $6^{th}$ allocation when it tries to follow China; although, for both targeted countries \emph{goods market efficiency} is most important.

These changes in policy priorities are not only dependent on which development footprints a country tries to track, but also on the country's initial conditions. For instance, Nigeria's top priority is health, regardless if it tries to copy Russia or Algeria. In contrast, the health pillar is not so important in Mexico's allocation profiles, and even less in Albania's where it is positioned at the bottom half. As we can see, policy priorities depend on the specific country--mode pair and, thus, on the network topology, budget and governance indicators of the imitator. This highlights the importance of considering the individual context of nations when providing policy advice on how to achieve development goals.

Finally, we would like to provide a global view of the development footprints of all countries in our data set (except countries in cluster 1 because they cannot follow anyone else). For this, we estimate the allocation profile of every country when trying to mimic each nation from the cluster above. Any given country has different development modes that it can follow. Then, the natural question is, which one should it choose? The answer is, obviously, context-specific and depends on defining a criterion that is relevant to the country. For example, a country may want to adopt a specific development mode because it is `easier' to achieve; because the end product would yield higher indicators; or because it puts more weight on developing a particular policy issue. Whichever the criterion is, the recommended priorities will depend on the network topology and the learning process of the country's agents. 

To illustrate this, we assume that the most feasible development mode is the one that requires the least changes to the allocation profile already adopted by a country (the one estimated in section \ref{sec:estimate_policy}). That is, if country $x$ has several options of countries to imitate, it should choose the one that requires the most similar allocation profile to the one already estimated for the 2006--2016 period. Here, we measure similarity between average allocation profiles through the matrix version of the Jaccard index. Common sense would tell us that similarity between allocation profiles is heavily determined by the proximity between the $x$'s and $y$'s development indicators. However, as we will show, similarities with the target's indicators are not the only driver. The game's learning dynamics and the network topology may be more important. In fact, most of the times, the most feasible target of $x$ is different from the the country with which $x$ has the highest target-indicator similarity (see \ref{app:triv_dev}). 

\begin{figure}[!hb]
    \centering
    \includegraphics[scale=.35]{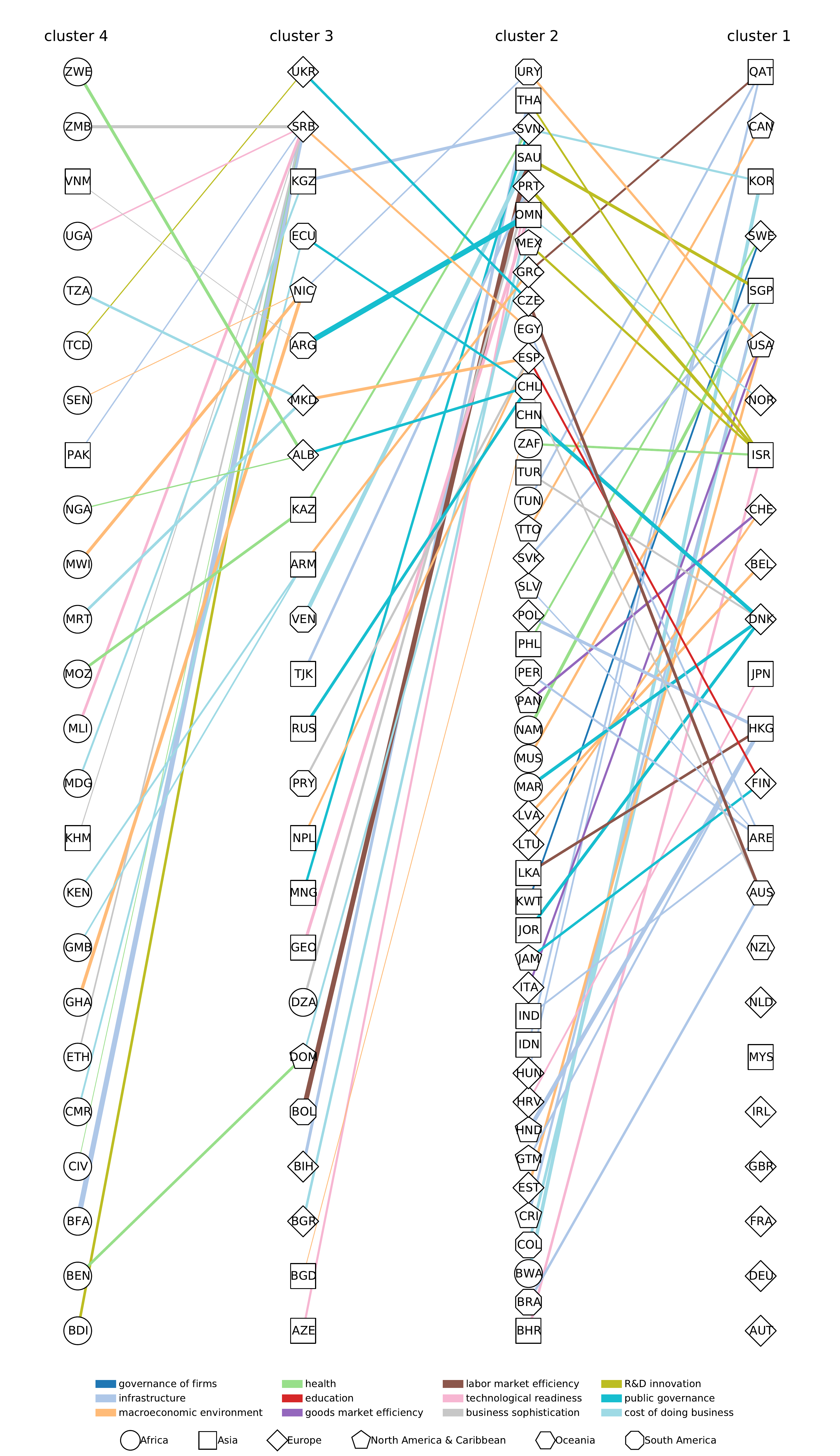}
    \caption{Map of development footprints. Edge thickness denotes the level of the Jaccard index. Number of simulations: 1,000 per pair of countries.}
    \centering\label{fig:followers}
\end{figure}

Figure \ref{fig:followers} provides an overall picture of countries and their development modes. Here, countries are arranged by cluster, and a connection between $x$ (from the lower cluster) and $y$ (from the higher cluster) means that $y$ is the most feasible development mode that $x$ can adopt (at least from $y$'s cluster). The weight of the links represents differences in feasibility (thicker is more feasible). Likewise, we have colored the edges in order to denote the top policy priority of the average allocation profile.

There are some patterns that stand out in Figure \ref{fig:followers}. For example, the catch-up processes from countries in cluster 4 are dominated by the \emph{health} and the \emph{cost of doing business} pillars. In cluster 3, the top priorities are more inclined toward \emph{public governance}, \emph{infrastructure} and the \emph{macro environment}. Members of cluster 2 exhibit more heterogeneity in their top priorities, where \emph{infrastructure} is the most frequent; while \emph{public governance}, \emph{R\&D} and the \emph{macroeconomic environment} are also common.

In terms of development modes, Figure \ref{fig:followers} suggests that there are countries with more `followable' footprints. For example, from cluster 3, Serbia concentrates more incoming edges from cluster 4.  In cluster 2, Slovenia is the leader. Finally, from cluster 1, the main mode is Israel, yet United States, Singapore, Hong Kong, United Arab Emirates and Denmark are also attractive targets. From these results, it is clear that, as countries develop, more options are available as feasible targets.

The likelihood of copying the transformative capabilities of a more advanced country not only depends on how large the required changes are, it also has to do with political economy considerations, aside from those contemplated in this study; for example, with how rapidly policy indicators can be tracked; with the resemblance between the productive structures of a country and its development mode; and with the `best practices' that a society might be allowed to pursue in the future. Nonetheless, the previous statistical exercise is very helpful to show that some development modes are more reachable than others and, thus, governments should be aware of this fact when setting targets for their indicators. In other words, although many countries want to follow the German mode, this European country might not be an easy catch.

\section{Conclusions\label{sec:conclusions}}

In this paper, we have investigated the challenging problem of how governments set policy priorities for achieving transformative development goals. We elaborated a theoretical framework that, in contrast with existing approaches, allows us to account for the complex network of interdependencies between policy issues, which can be highly specific to the context of each nation. Through a political economy game on a network, we model how governments adapt policy priorities in order to discourage decentralized forms of corruption while attaining specific targets. Our computational analysis of the model shows that it can produce well-known stylized facts about corruption and socioeconomic performance, as well as replicating corruption levels across countries. 

Besides providing evidence of external validation, we subject the model to multiple internal validation tests, demonstrating the relevance of specific social mechanisms. In particular, these sensitivity tests show the importance of the government's allocation heuristic, the relevance of co-evolutionary learning among agents and the spillover effects created by the network of policy issues. That is, we show that the causal channels assumed in the model are informative, since they explain the model's outputs between (\emph{e.g.}, corruption levels) and within (\emph{e.g.}, allocation and contribution profiles) countries. Hence, we argue that the theoretical support of our methodology is well grounded and, thus, our policy prescriptions are empirically sound and provide an interpretation that is consistent with the models' logic.  

Using data on 79 development indicators for 117 countries, we provide a first application of this framework. Our computational procedure is rather simple since only one parameter is estimated. The empirical analysis reveals that, in the sampling period of 2006--2016, countries pursued a wide variety of development modes reflected in their policy priorities. This retrospective analysis indicates that the observed priorities are not closely related to the empirical gaps between development indicators, as a simple rule for the allocation of budgetary resources would suggest. Moreover, the simulation results of the ABM indicate that context matters when designing policy guidelines to produce a structural change in any economy. 

In the prospective application, we argue that there are different development modes that any country can follow in order to generate transformative capabilities.  Because each of these modes has priority rankings, the relevant aspect of pursuing a particular strategy has to do with the consistency of its policies; that is, with following closely the suggested prioritization scheme. We find, for example, that Albania can follow a Spanish mode that emphasizes goods markets efficiency, financial market development, public governance and infrastructure; or undertake a Chinese mode whose main four concerns are goods market efficiency, R\&D innovation, labor market efficiency and the macroeconomic environment; among other possible development modes. 

An empirical extension of this paper is to design and analyze other criteria for establishing which countries can be attractive targets. One possibility is to specify a sample of more advanced countries with an appealing GDP per capita or with a good-enough level of the human development index. Then, this set can be further constrained in terms of similarity in its productive structure, international trade composition, culture, geographical proximity, political regime or any combination of the above. Likewise, for a further analysis of the robustness of our framework, it can be implemented with different data, such as development indicators at a sub-national level, and then tested for its capabilities of replication. It would be also enlightening to explore different methods of network calibration and alternative lag structures of the variables involved. These exercises could be beneficial for improving the model's goodness of fit and for increasing the number of replicated outcomes. 

Although, we did not compute marginal effects for each public issue in this paper; this type of empirical results could also be obtained from  our methodology. Thus, another interesting extension would be to remove investments in certain policy issues by means of sensitivity analyses. That is, in the government’s behavioral model, we can preclude allocations to a specific issue, while the associated indicator can still grow due to spillover effects. Through this procedure, one could measure relative impacts on a counter-factual basis and using different dependent variables such as convergence time and corruption levels.

\bibliography{references2}

\newpage
\appendix

\onecolumn
\captionsetup{font=scriptsize}

\newpage

\section{Descriptive statistics}\label{app:descriptive}

Table \ref{tab:list_countries} provides a list of all clusters and a list of all countries in each cluster. Table \ref{tab:desc_stats} presents summary statistics for each indicator in the study.

\begin{table}[!htb]
   \linespread{0.95} \centering\footnotesize
    \caption{List of countries by cluster}\label{tab:list_countries}
    \begin{tabular}{p{3cm}cp{8cm}}
        \hline
        \textbf{Cluster} & \textbf{Number of countries} & \textbf{Countries}\\\hline
        1) High  & 24 &
        ARE AUS AUT BEL CAN CHE DEU DNK FIN FRA GBR HKG IRL ISR JPN KOR MYS NLD NOR NZL QAT SGP SWE USA \\\hline
        2) Mid-high  & 45 &
        BHR BRA BWA CHL CHN COL CRI CZE EGY ESP EST GRC GTM HND HRV HUN IDN IND ITA JAM JOR KWT LKA LTU LVA MAR MEX MUS NAM OMN PAN PER PHL POL PRT SAU SLV SVK SVN THA TTO TUN TUR URY ZAF \\\hline
        3) Mid-low  & 24 &
        ALB ARG ARM AZE BGD BGR BIH BOL DOM DZA ECU GEO KAZ KGZ MKD MNG NIC NPL PRY RUS SRB TJK UKR VEN \\\hline
        4) Low  & 24 &
        BDI BEN BFA CIV CMR ETH GHA GMB KEN KHM MDG MLI MOZ MRT MWI NGA PAK SEN TCD TZA UGA VNM ZMB ZWE \\\hline
    \end{tabular}
\end{table}

\newpage

\begin{spacing}{.8}
\footnotesize
\begin{longtable}{lp{9cm}ccrr}
   \caption{Descriptive statistics}\label{tab:desc_stats}\\
    \toprule
    \multicolumn{2}{l}{{Pillars and indicators}} & {N2} & {Switch} & {Mean} & {SD}\\\midrule
    \endfirsthead
    \caption[]{\emph{continued}}\\
    \toprule
    \multicolumn{2}{l}{{Pillars and indicators}} & {N2} & {Switch} & {Mean} & {SD}\\\midrule
    \endhead
    \bottomrule
    \multicolumn{6}{p{0.8\textwidth}}{\scriptsize{Notes:} The column \textit{N2} indicates whether we applied the skewness correction and column \textit{Switch} informs about indicators that have been inverted. All statistics are based on the full sample.}
    \endlastfoot
    \input{desc_stat}

\end{longtable}
\end{spacing}

\normalsize
\newpage

\section{Development indicators and network calibration}\label{app:plots}

\subsection{Development indicators across clusters}

\begin{figure}[h!]
    \centering
    \includegraphics[scale=.7]{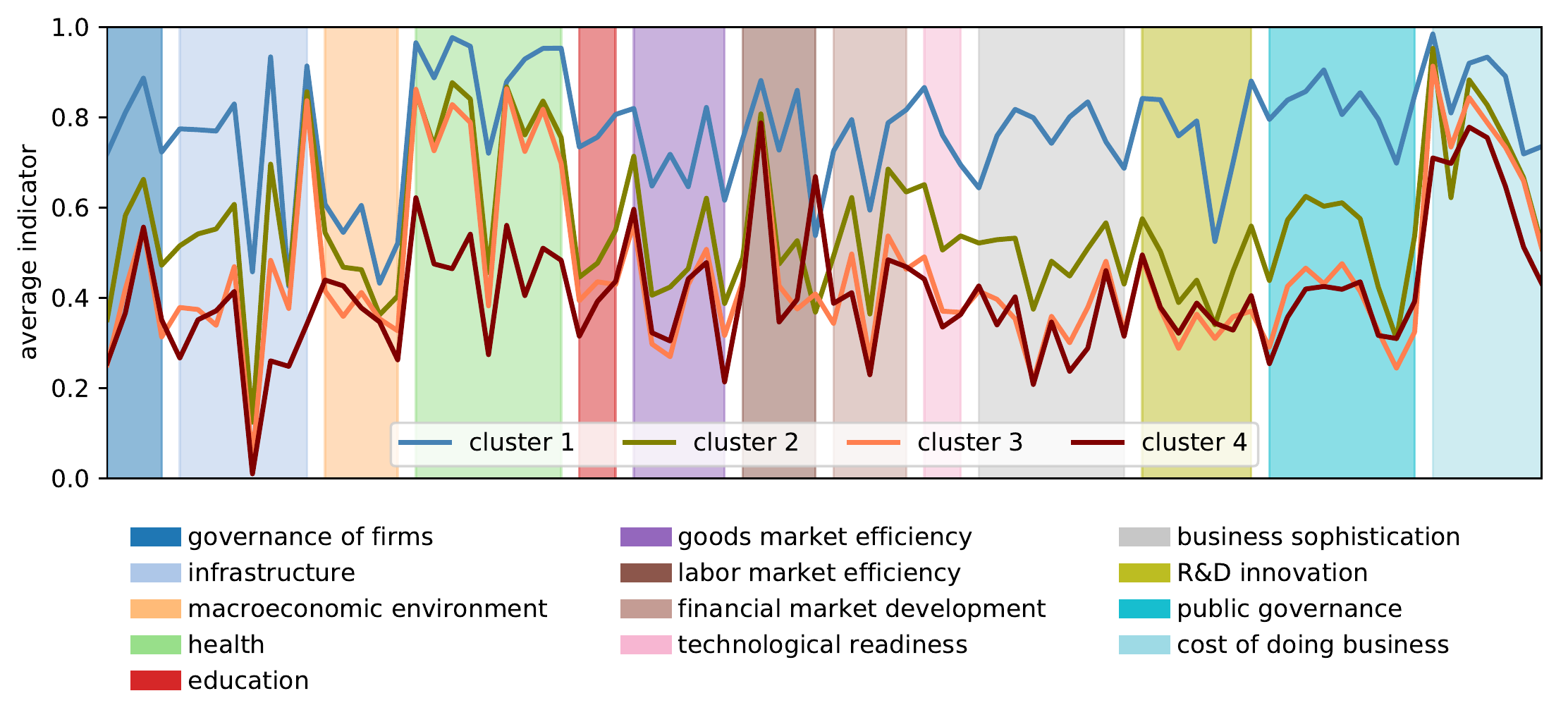}
    \caption{Each color shade corresponds to a development pillar. The wider the shaded area, the more development indicators in the corresponding pillar. For countries without observations for 2016, we take the most recent one. The clusters correspond to those in Table \ref{tab:list_countries}.}
    \centering\label{fig:indicators_series}
\end{figure}

\subsection{Network estimation for individual countries}\label{app:jaccard}

With the methodology for estimating directed networks, it is possible to estimate the interdependencies among policy issues for any country in the data. Once this is done, we can analyze how similar are the topologies of all pair of countries. For this, we employ the weighted Jaccard index

\begin{equation}
    S(\mathbb{A}, \mathbb{B}) = \frac{\sum_i \sum_j \min(\mathbb{A}_{ij}, \mathbb{B}_{ij})}{\sum_i \sum_j \max(\mathbb{A}_{ij}, \mathbb{B}_{ij})},\label{eq:jaccard}
\end{equation}
which measures the degree of similarity between the two directed weighted networks encoded in adjacency matrices $\mathbb{A}$ and $\mathbb{B}$. The range of $S$ goes from zero (if the networks are entirely different) to one (if $\mathbb{A}=\mathbb{B}$). This version of the Jaccard index must not be confused with the membership index used in section \ref{sec:priorities}, which compares similarity between two sets.

\begin{figure}[h!]
    \centering
    \includegraphics[scale=.5]{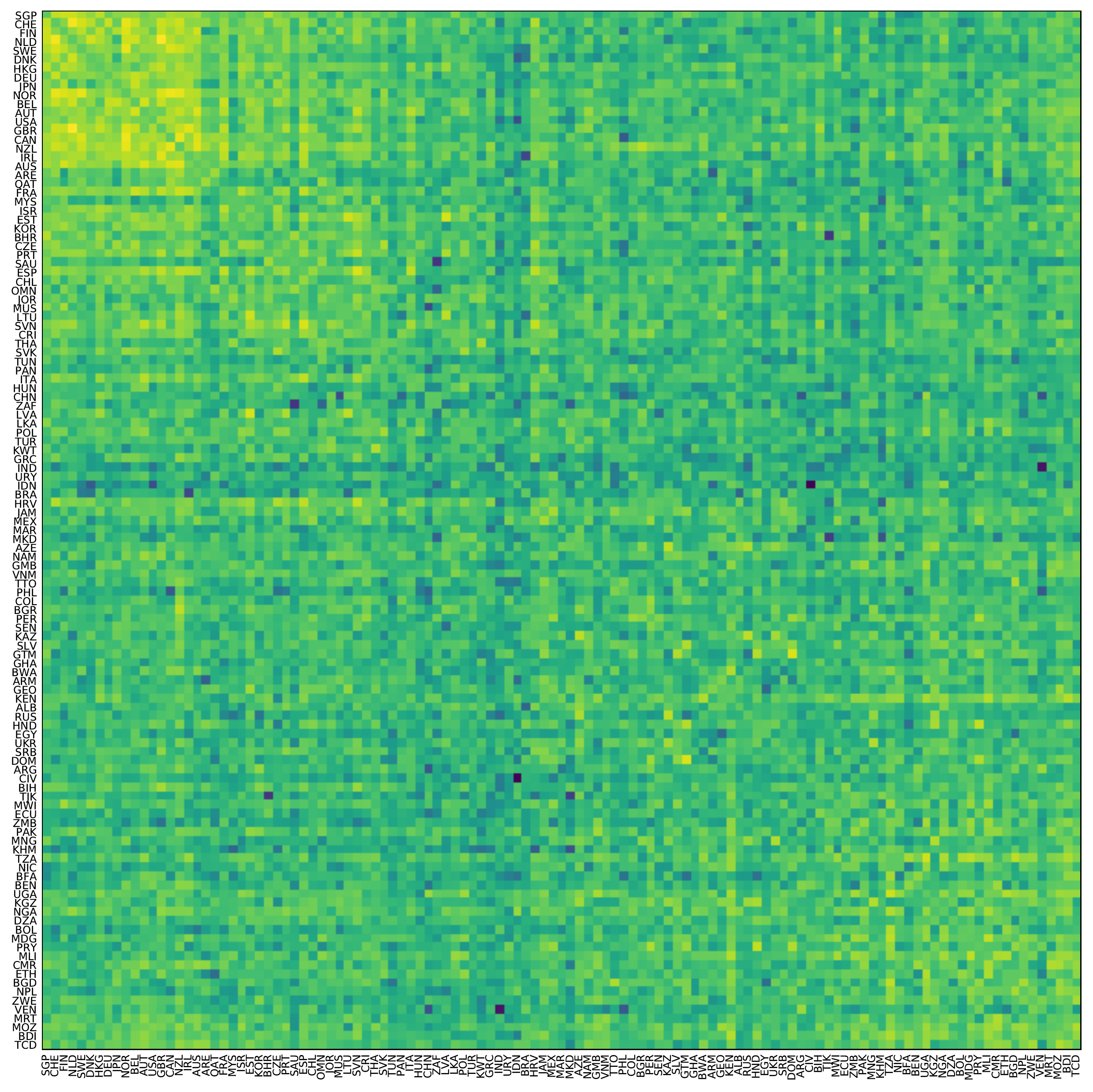}
    \caption{Network similarity between countries. Yellow means higher Jaccard indices, while blue means lower ones.}
    \centering\label{fig:jaccard}
\end{figure}

Figure \ref{fig:jaccard} shows a matrix of Jaccard indices between every pair of countries. We have sorted the rows and columns according to the average indicators, as computed in equation \ref{eq:data_develop}, which is indicative of how developed a country is. Then, entries concentrated around the diagonal would reveal that countries in similar stages of development have some degree of similarity in their networks of indicators. From the figure, it is clear that this is the case for countries with high average indicators. In addition, it is possible to identify two more clusters around the diagonal: one for middle-high income countries, and another for the least developed countries. Despite these patterns, several countries share relatively similar topologies with nations that are in a different development stage.

\newpage

\section{Model calibration}\label{app:calibration}

In principle, the model can be used to estimate the allocation profile of a specific country without calibrating free parameters. However, in order to perform a cross country analysis with aggregated outputs, it is convenient to calibrate the model. In this case, we can obtain corruption estimates similar to the ones observed in real-world data. This is, precisely, the purpose of introducing parameter $\gamma$ in equation \ref{eq:indicator}.\footnote{For a single country, the order of policy priorities does not change significantly across different levels of $\gamma$, although their differences do, which might affect aggregation. The Spearman correlation coefficient comparing the rank-order correlation between profiles corresponding to different $\gamma$s within the same country yields more than 0.9 in all countries. For this reason, it is safe to assume $\gamma=1$ when making country-level inference about the order of policy priorities.}

The calibration of $\gamma$ across countries can be thought as estimating parameter heterogeneity in a regression model. However, the estimation method differs significantly due to the computational nature of the model. More specifically, estimating $\gamma$ corresponds to solving the classification problem of clustering. In other words, calibrating $\gamma$ translates into finding a set of distinct values under which all countries can be classified. This set should be such that we minimize the error between the observed and the estimated levels of corruption while penalizing potential overfitting. On one hand, the model without calibration implies a homogeneous $\gamma$, which means that there is only one cluster. On the other, a unique $\gamma$ for each country implies a cluster for each nation, which is an obvious overfit. Note that even an overfit does not guarantee a perfect fit. This is so because, for a specific country, $\gamma$ lives in $(0,\theta)$. If $\gamma=0$, then the indicators do not grow and the model never converges. The other extreme happens when $\gamma$ is too large. In this case, the growth of the indicator in a single step may be too large to reach convergence. Therefore, $\gamma$ itself may not be enough to provide the necessary variation for a perfect fit.

Since the units of the empirical indicator of \emph{diversion of public funds} are different from the ones of the theoretical variable of corruption, as reflected by equation \ref{eq:data_corruption}, we aim to fit the model exploiting the empirical structure of relative differences in corruption between countries. To achieve this, our calibration strategy consists of two methods. The first one finds the optimal reference point to compute relative levels of corruption. The second determines the distinct values that $\gamma$ should take in order to minimize the mean squared error (MSE) and control for overfitting.

\subsection{The `ratios' method}

This step begins with a proposed set $\Gamma = \{ \gamma_1 \dots \gamma_n \}$ of distinct values. The idea is to find the best way to classify countries across $\Gamma$ such that the MSE between the empirical indicator of \emph{diversion of public funds} and $\bar{D}$ in equation equation \ref{eq:data_corruption} is minimum. In order to achieve this, we perform the following algorithm.

\begin{enumerate}
    \item Start with a reference country $r$.
    \item Pick a $\gamma \in \Gamma$ and compute $\bar{D}$ for the reference country.
    \item Select an evaluation country $e$ and compute the rate $I_e/I_r$, where $I$ is the empirical indicator of \emph{diversion of public funds}.
    \item Choose a $\gamma \in \Gamma$ for the evaluation country such that $|I_e/I_r - \bar{D}_e/\bar{D}_r|$ is minimal.
    \item Repeat steps 3 and 4 until all countries have been evaluated.
    \item Compute the mean squared error $\frac{1}{S}\sum_{i=1}^S (I_e/I_r - \hat{D}_e/\hat{D}_r)^2$, where $S$ is the number of countries in the sample.
    \item Repeat steps 2 to 6 for every $\gamma \in \Gamma$ and choose the reference $\gamma$ that minimizes the MSE.
    \item Repeat steps 1 to 6 to obtain the reference country that minimizes the MSE.
\end{enumerate}

By the end of this procedure, we obtain the different values of $\gamma$ that minimize the MSE. These values can be a subset of $\Gamma$, so this step can preclude overfitting. Of course, overfitting depends on the resolution of the proposed set $\Gamma$. The larger this set is, the bigger the computational burden. Here, our strategy is to initially compute the model for every country and every $\gamma \in [1, 30]$ with a resolution of 117, which is the number of countries in the sample. The upper bound has been determined by trial an error, after noticing that, for larger $\gamma$s, the model is unlikely to converge. Hence, for clarity of exposition, let us denote the set of $\gamma$s resulting from this method as $\Gamma^*$.

\subsection{The `jump' method}

This method tries to further reduce the number of distinct values of $\gamma$, in order to minimize overfitting. This is a classic problem in the clustering literature, also known as `finding the number of clusters in a data set'. There are many methods designed for this purpose, and most of them build on the principle of minimizing the distances within clusters and maximizing the distances between them (like the Ward's method used in section \ref{sec:data}). By choosing a subset of $\Gamma^*$, we seek to minimize the single-dimensional distance within clusters, \emph{i.e.} to minimize the MSE. For the reason, we use the `jump' method developed by \cite{sugar_finding_2003}.

The jump method builds on information theory and, in particular, on the engineering problem of optimally coding a continuous signal into a discrete set of symbols. The analogy here is that the signal is the continuous line $[1, 30]$ which, if estimated with enough resolution, would yield 117 unique values for $\gamma$; a clear overfit equivalent to assigning one cluster to each observation. The idea of the jump method is to find a subset $\Gamma^J \subset \Gamma^*$ such that the subsequent marginal contributions of increasing the subset become relatively small. \cite{sugar_finding_2003} show that the true number of clusters relates to the inverse of the root mean squared error (RMSE). Hence, by computing $\text{MSE}^{-1/2}$ for all sizes of $\Gamma^*$, we can estimate the number of unique values of $\gamma$.

There is an additional challenge in this context. A subset $\Gamma^J \subset \Gamma^*$ can be arbitrarily defined, for example, if $|\Gamma^*|=100$ and $|\Gamma^J|=10$ there are $100\choose10$ ways to determine $\Gamma^J$. For this reason, we sample the space of possible subsets and obtain the best ones by following the next algorithm. 

\begin{enumerate}
    \item Propose a random set $\Gamma' \subseteq \Gamma$.
    \item Obtain $\Gamma^* \subseteq \Gamma'$ and its size $h$.
    \item If the MSE of $\Gamma^*$ is smaller than the ones recorded so far for sets of size $h$, then $\Gamma^*$ is the best set of size $h$ found so far.
    \item Repeat steps 1 to 3 until each one of the sets of sizes $1, \dots, |\Gamma|$ have been sampled enough.
    \item Once with the best set $\Gamma^*$ of each size, compute their corresponding mean squared errors $\{ \text{MSE}_1, \dots, \text{MSE}_H\}$.
    \item Pick the set $\Gamma^*$ such that its size $h^* = \text{argmax} \{ \text{MSE}^{-1/2}_h - \text{MSE}^{-1/2}_{h-1} \}_1^H$.
\end{enumerate}

\begin{figure}[!htb]
    \centering
    \includegraphics[width=0.48\textwidth]{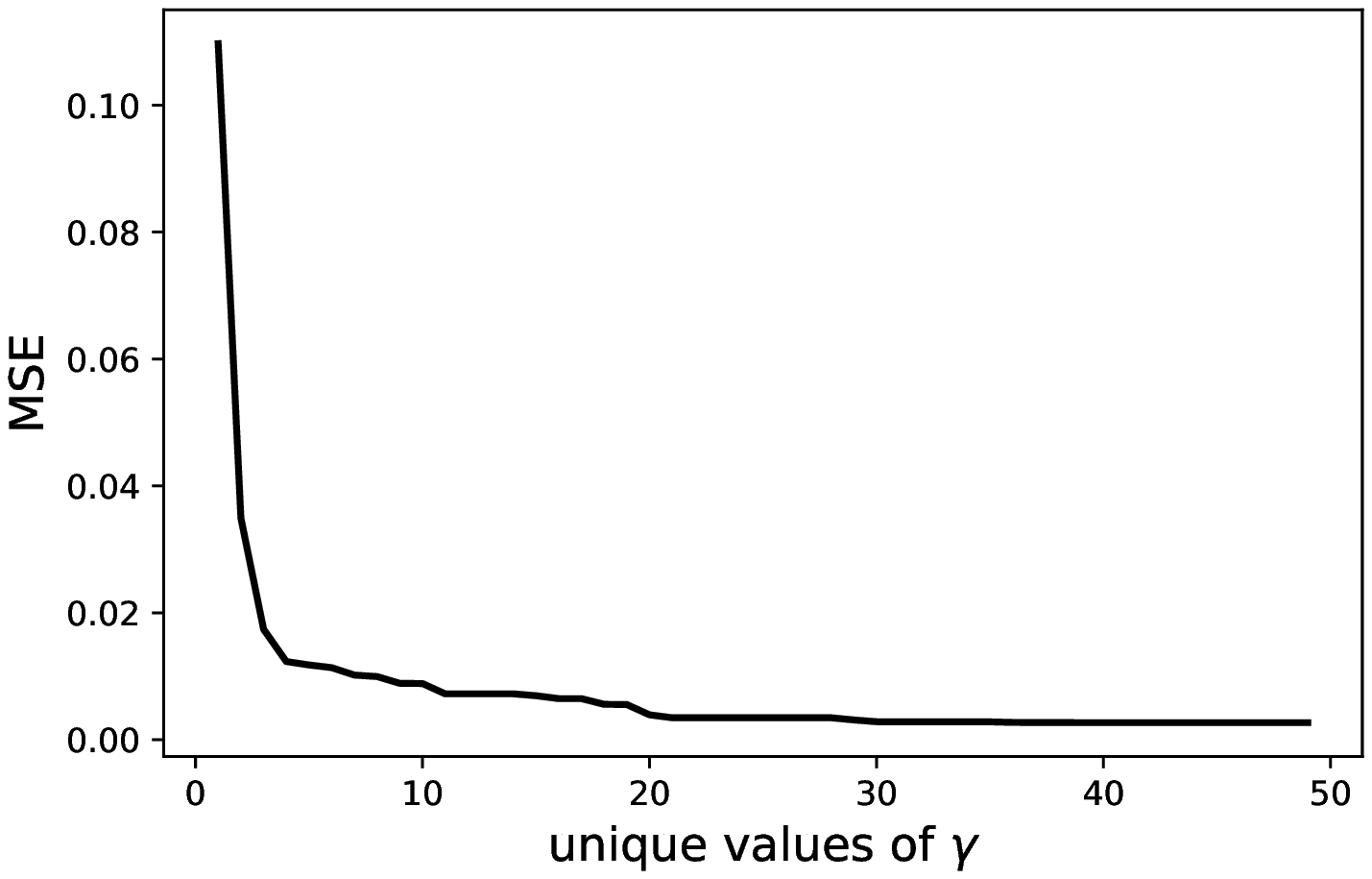}\\
    \includegraphics[width=0.48\textwidth]{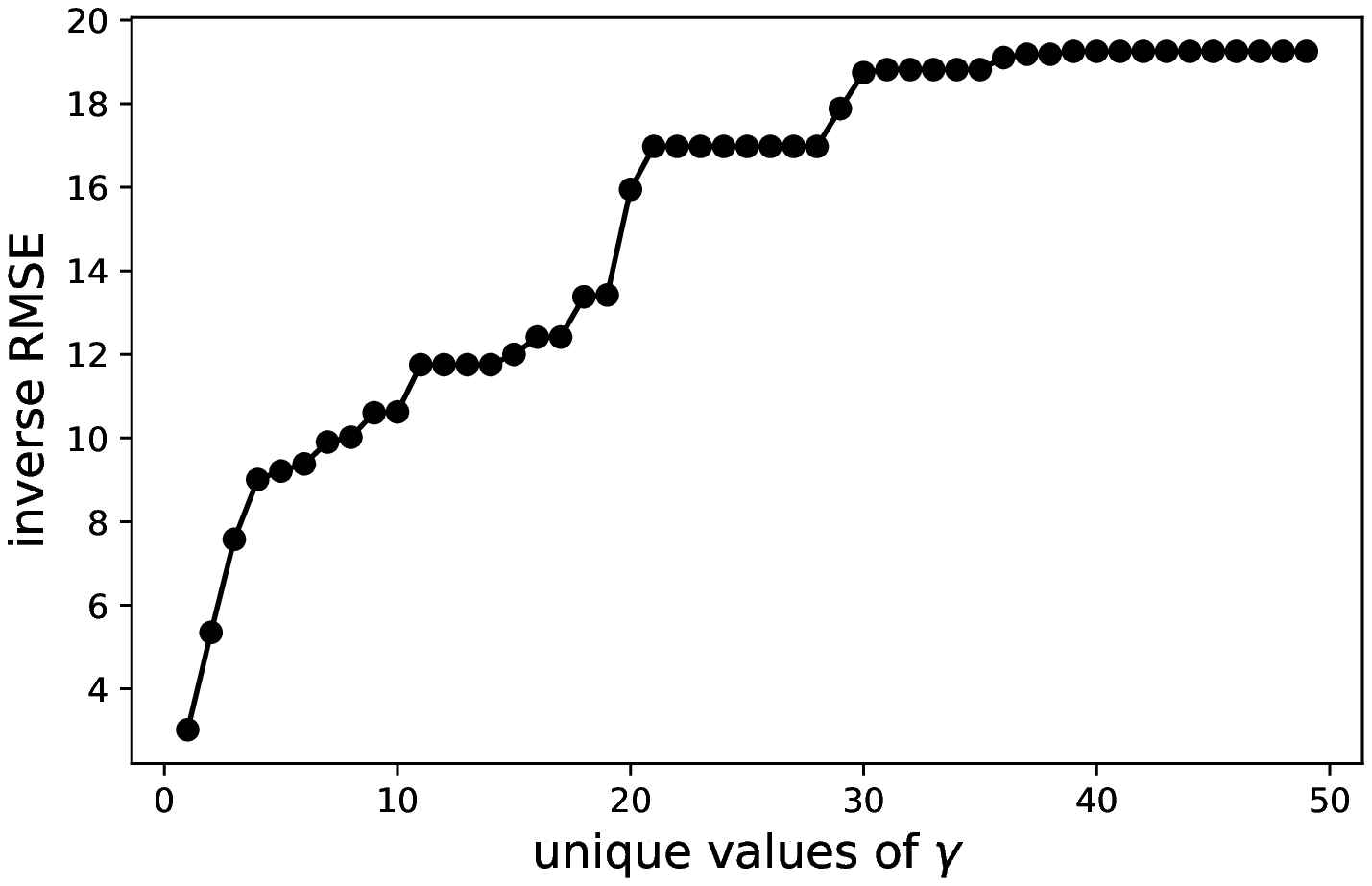}
    \includegraphics[width=0.48\textwidth]{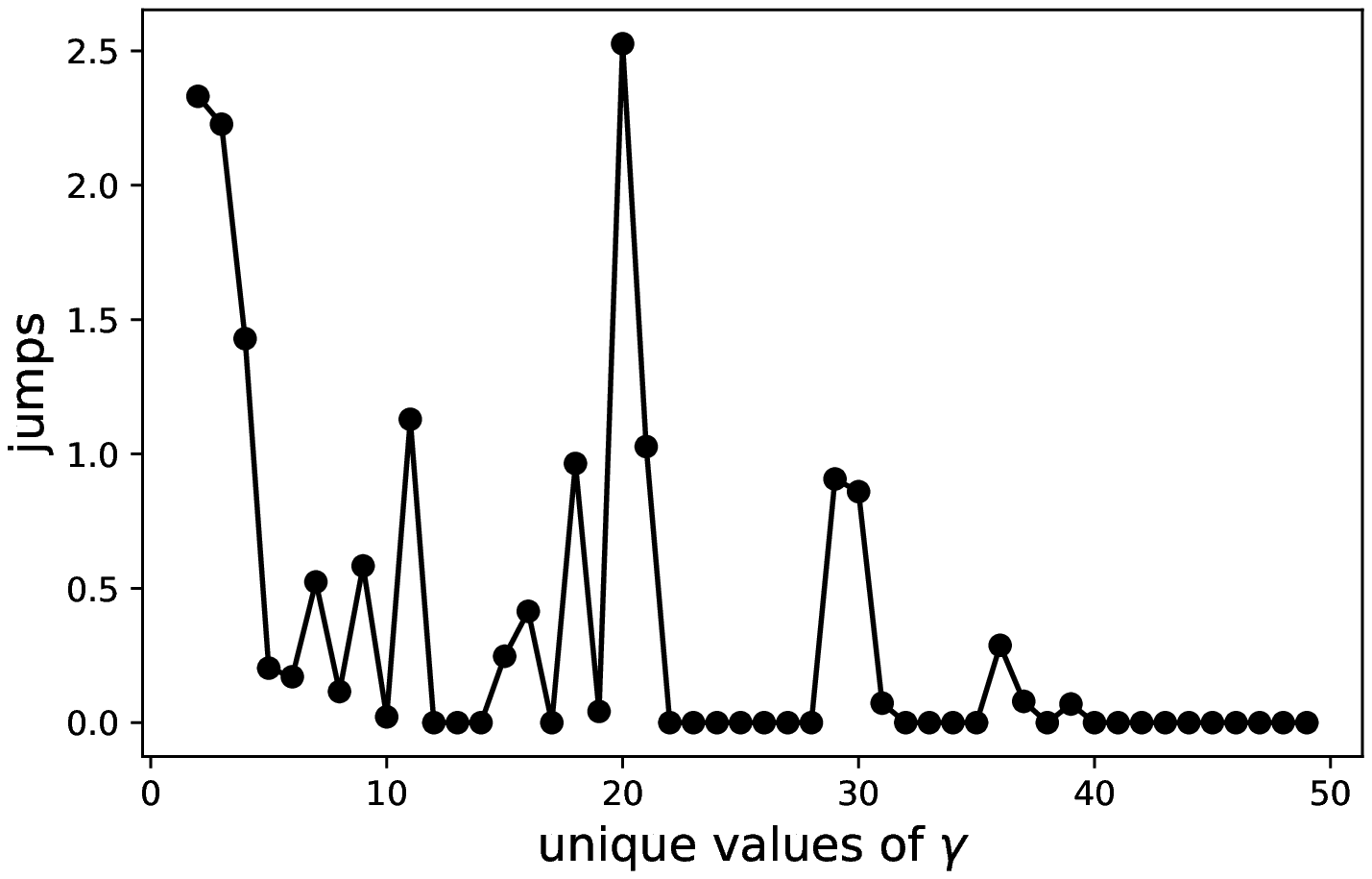}
    \caption{Model calibration. Estimations were obtained from over 1 million samples.}
    \centering\label{fig:growthCorrelation}
\end{figure}

The previous algorithm yields the different values of $\gamma$ that balance error and overfit minimization. To illustrate the outcomes of the entire estimation strategy, Figure \ref{fig:growthCorrelation} shows the MSE reduction with more values for $\gamma$ (top panel), the growth of the inverse RMSE (bottom left panel) and the biggest jump at 20 distinct values (bottom right panel).

\newpage

\section{Additional sensitivity tests}\label{app:sensitivity}

\subsection{Sensitivity of corruption}\label{app:corrNoNet}

\begin{figure}[h!]
    \centering
    \includegraphics[scale=.5]{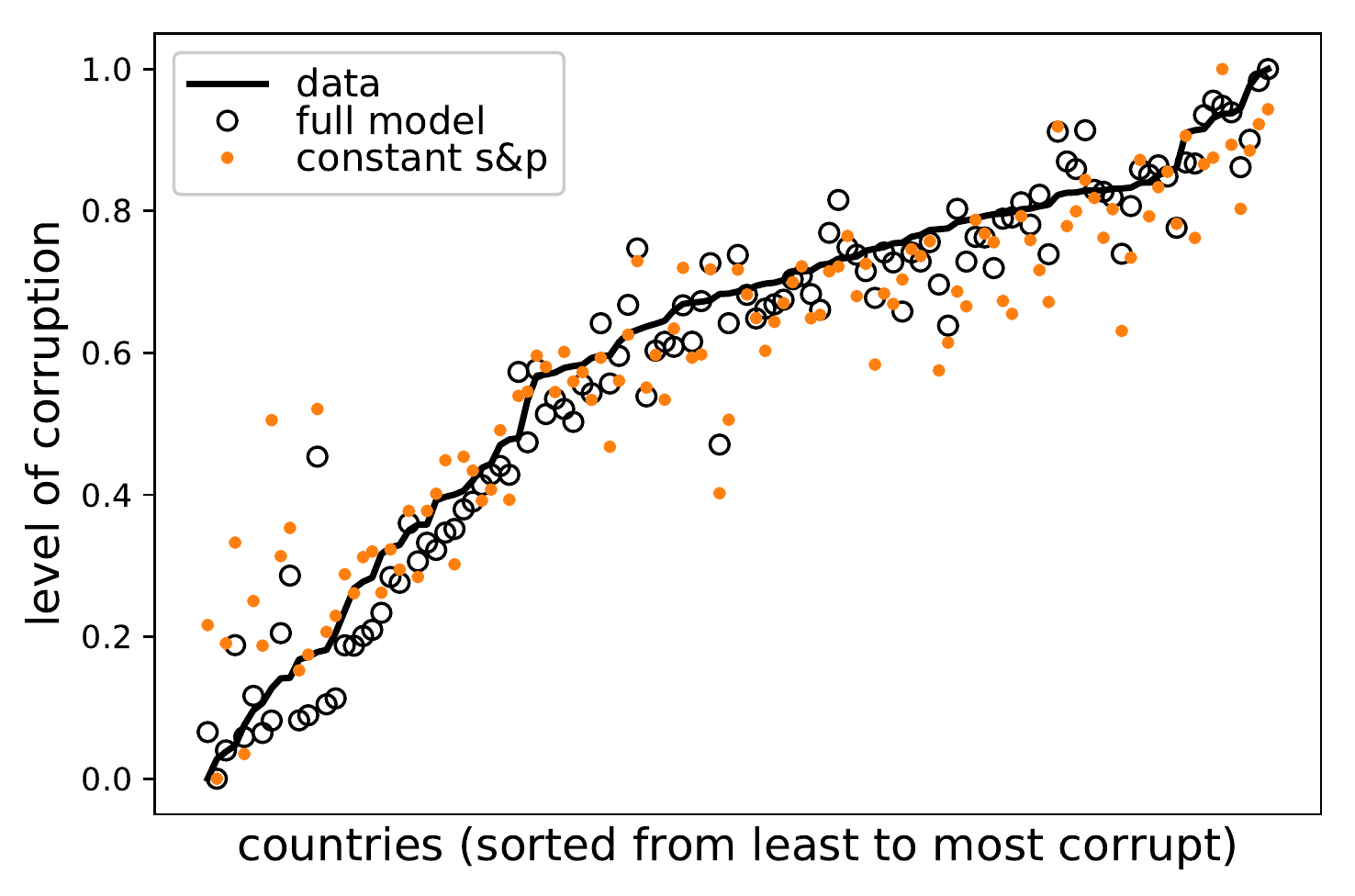}
    \includegraphics[scale=.5]{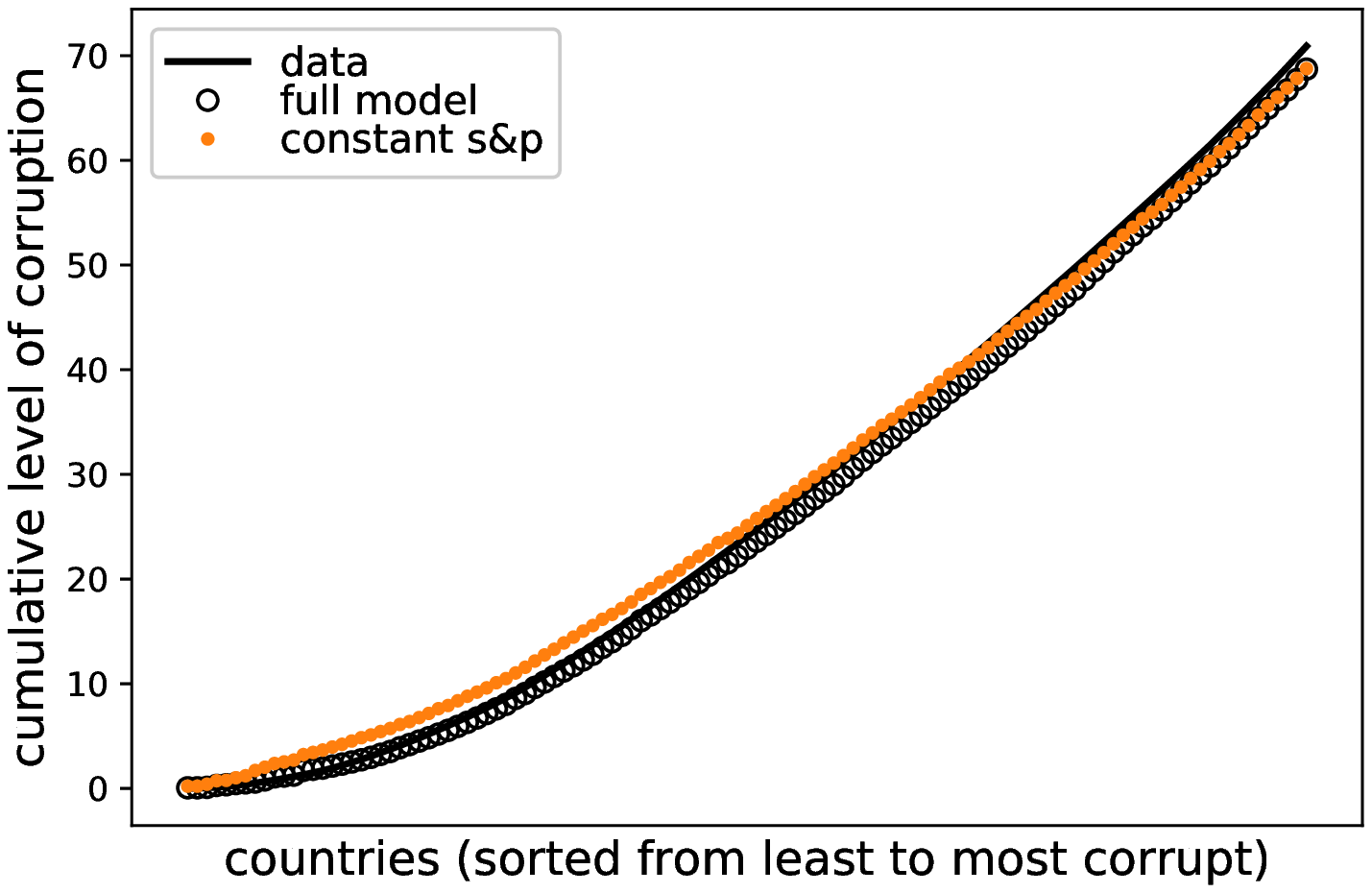}
    \caption{The role of supervision and punishment. Model with fixed values $f_{R,t}=f_{C,t}=0.5$ for every country and. Left panel: point-estimate comparison. Right panel: accumulated values of corruption across countries.}
    \centering\label{fig:sensi3c}
\end{figure}

Figure \ref{fig:sensi3c} shows the effect of fixing $f_{R,t}$ and $f_{C,t}$ at a constant value ($0.5$) across countries and in all periods. The biases with respect to the full model (and the data) are particularly notorious among low and high corruption countries, as shown in the point-estimates presented in the left panel. The magnitude of these biases, or course, varies with the value to which $f_{R,t}$ and $f_{C,t}$ are fixed. In our experience, 0.5 is a conservative specification.

\begin{figure}[h!]
    \centering
    \includegraphics[scale=.5]{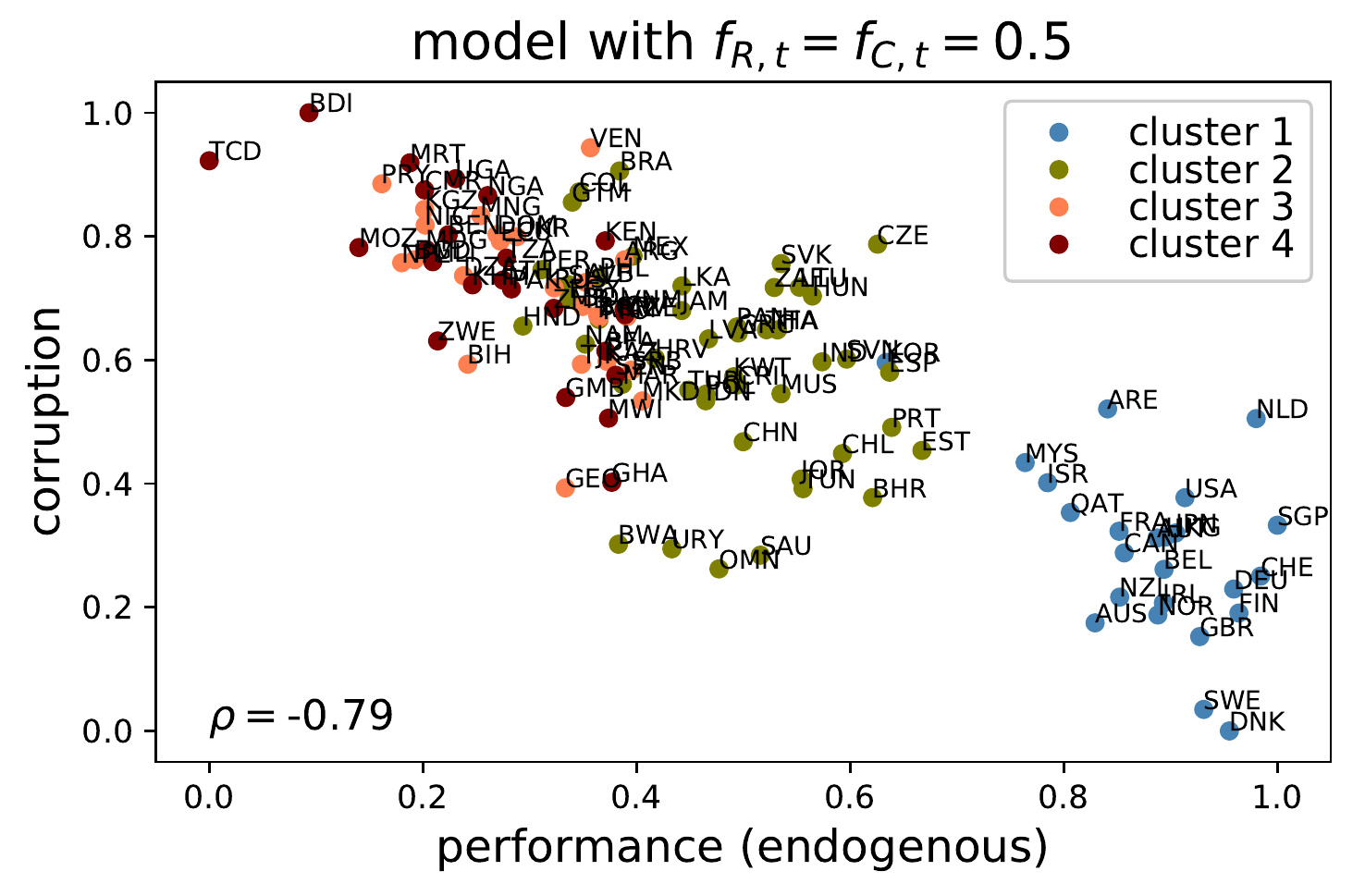}
    \includegraphics[scale=.5]{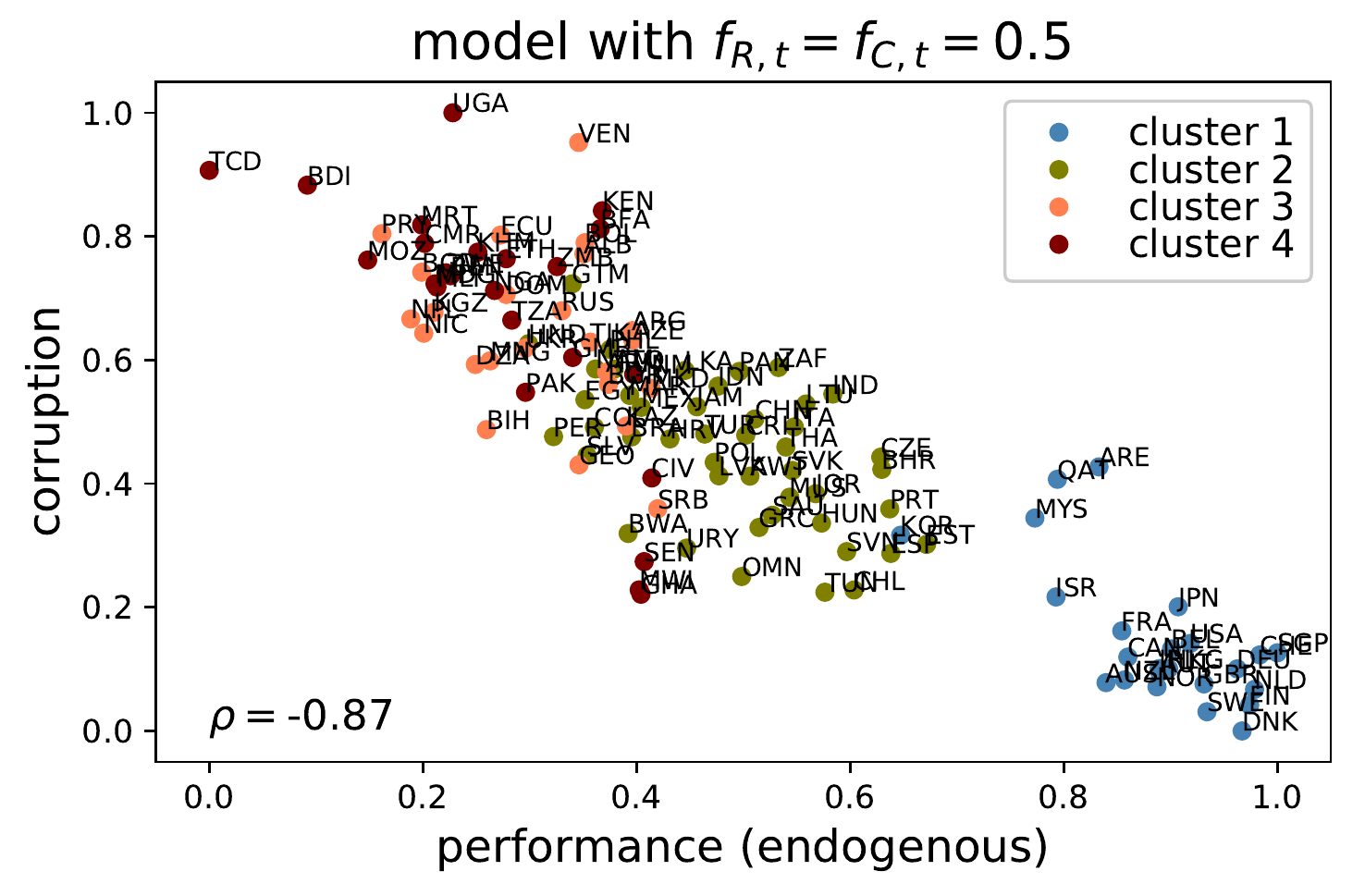}
    \caption{Additional tests for internal validation. Left panel: The corruption-performance relationship with fixed $f_{R,t}$ and $f_{C,t}$.  Right panel: The corruption-performance relationship without network spillovers.}
    \centering\label{fig:sensi3b}
\end{figure}

Additionally, the left panel in Figure \ref{fig:sensi3b} shows that fixing $f_{R,t}$ and $f_{C,t}$ distorts slightly the corruption-performance relationship. Notice that, in contrast to the empirical data, several countries in cluster 3 and 4 have corruption levels similar or below those simulated for some countries in cluster 1. Thus, these artificial data are unable to neatly replicate the absence of an overlapping in corruption levels suggested by the empirical data (see the left panel of Figure \ref{fig:corruption}). Likewise, the correlation coefficient goes down 6 points with respect to the full model.  Consequently, these two variables are not the only factors that help to explain the observed stylized facts.

The right panel in Figure \ref{fig:sensi3b} shows that the effect from removing spillovers on the corruption-performance relationship is more modest. This, however, does not invalidate the spillovers as an important mechanism. it only suggests that, it is not relevant to explain this particular stylized fact.

\newpage\section{Non-Trivial Development Footprints}\label{app:triv_dev}

\rowcolors{2}{gray!25}{white}
\begin{table}[!htb]
\tiny
    \centering
    \caption{Countries, most feasible targets and trivial targets. Most feasible targets are those with the highest Jaccard index between the estimated allocation profile and the one needed to reach the target country. Trivial targets are the ones with the highest target-indicator similarity.}\label{tab:triv_dev}
    \begin{tabular}{ccccccccccc}
        \toprule
        Follower & Target & Trivial Target & & Follower & Target & Trivial Target & & Follower & Target & Trivial Target\\
        \midrule
ALB & CHL & HND && GRC & QAT & QAT && NPL & ESP & GTM\\
ARG & OMN & MEX && GTM & HKG & QAT && OMN & NOR & QAT\\
ARM & GRC & MAR && HND & HKG & QAT && PAK & SRB & AZE\\
AZE & PRT & MAR && HRV & JPN & QAT && PAN & CHE & QAT\\
BDI & SRB & NIC && HUN & QAT & QAT && PER & ARE & QAT\\
BEN & DOM & NIC && IDN & QAT & QAT && PHL & SWE & QAT\\
BFA & SRB & NPL && IND & ARE & MYS && POL & HKG & QAT\\
BGD & CHN & GTM && ITA & USA & KOR && PRT & ISR & QAT\\
BGR & OMN & HRV && JAM & FIN & QAT && PRY & ESP & HND\\
BHR & ISR & QAT && JOR & DNK & QAT && RUS & CHL & CHN\\
BIH & THA & HRV && KAZ & SVN & LVA && SAU & SGP & QAT\\
BOL & SAU & HND && KEN & ARM & KAZ && SEN & NIC & SRB\\
BRA & AUS & QAT && KGZ & SVN & HND && SLV & ARE & QAT\\
BWA & KOR & QAT && KHM & SRB & TJK && SRB & EGY & OMN\\
CHL & AUS & KOR && KWT & SWE & QAT && SVK & SGP & QAT\\
CHN & DNK & QAT && LKA & HKG & QAT && SVN & KOR & KOR\\
CIV & SRB & SRB && LTU & CHE & KOR && TCD & UKR & BOL\\
CMR & ECU & NIC && LVA & BEL & QAT && THA & ISR & ARE\\
COL & SGP & QAT && MAR & DNK & QAT && TJK & PRT & HND\\
CRI & SGP & QAT && MDG & KGZ & NIC && TTO & CAN & QAT\\
CZE & AUS & KOR && MEX & ISR & QAT && TUN & QAT & QAT\\
DOM & MEX & GTM && MKD & ESP & MAR && TUR & DNK & KOR\\
DZA & PRT & MAR && MLI & SRB & NIC && TZA & MKD & SRB\\
ECU & CHL & HND && MNG & SVN & HND && UGA & SRB & NPL\\
EGY & ARE & QAT && MOZ & KAZ & BOL && UKR & CZE & HRV\\
ESP & FIN & KOR && MRT & MKD & TJK && URY & USA & QAT\\
EST & USA & MYS && MUS & USA & QAT && VEN & SAU & PER\\
ETH & SRB & BOL && MWI & NIC & SRB && VNM & ARG & AZE\\
GEO & OMN & MAR && NAM & SGP & QAT && ZAF & ISR & NZL\\
GHA & NIC & SRB && NGA & ALB & KAZ && ZMB & SRB & NPL\\
GMB & ARM & NIC && NIC & URY & HND && ZWE & ALB & VEN\\
        \bottomrule
    \end{tabular}
\end{table}

\end{document}

%% file: desc_stat.tex
\\
\multicolumn{2}{l}{\textbf{Governance of firms}}\\
P01\_1 & Ethical behavior of firms & No & No & 0.42 & 0.22 \\ 
P01\_2 & Strength of auditing and reporting standards & No & No & 0.55 & 0.19 \\ 
P01\_3 & Efficacy of corporate boards & No & No & 0.58 & 0.14 \\ 
P01\_4 & Protection of minority shareholders' interests & No & No & 0.51 & 0.19 \\ 

\\
\multicolumn{2}{l}{\textbf{Infrastructure}}\\
P02\_1 & Quality of overall infrastructure & No & No & 0.50 & 0.24 \\ 
P02\_2 & Quality of roads & No & No & 0.49 & 0.24 \\ 
P02\_4 & Quality of port infrastructure & No & No & 0.52 & 0.21 \\ 
P02\_5 & Quality of air transport infrastructure & No & No & 0.61 & 0.19 \\ 
P02\_6 & Available airline seat km/(week*population), millions  & Yes & No & 0.14 & 0.21 \\ 
P02\_7 & Quality of electricity supply & No & No & 0.60 & 0.27 \\ 
P02\_8 & Mobile cellular subscriptions (per 100 people)  & No & No & 0.31 & 0.13 \\ 
P02\_9 & Improved sanitation facilities, urban (\% of urban population with access) <norm2 & No & No & 0.77 & 0.28 \\ 

\\
\multicolumn{2}{l}{\textbf{Macroeconomic environment}}\\
P03\_1 & Inflation, annual \% change  & No & Yes & 0.42 & 0.17 \\ 
P03\_2 & General government debt, \% GDP  & Yes & No & 0.39 & 0.23 \\ 
P03\_3 & Foreign direct investment, net inflows (BoP, current USD)  & Yes & No & 0.47 & 0.19 \\ 
P03\_4 & Imports as a percentage of GDP  & Yes & Yes & 0.39 & 0.23 \\ 
P03\_5 & Exports as a percentage of GDP  & Yes & No & 0.41 & 0.23 \\ 

\\
\multicolumn{2}{l}{\textbf{Health}}\\
P04\_3 & Tuberculosis cases/100,000 pop.  & Yes & Yes & 0.79 & 0.27 \\ 
P04\_6 & Business impact of HIV/AIDS & No & No & 0.66 & 0.21 \\ 
P04\_7 & Infant mortality, deaths/1,000 live births  & No & Yes & 0.71 & 0.31 \\ 
P04\_8 & Adolescent fertility rate (births per 1,000 women ages 15-19)  & No & Yes & 0.77 & 0.20 \\ 
P04\_9 & Health expenditure, public (\% of GDP)  & No & No & 0.45 & 0.26 \\ 
P04\_10 & Immunization, DPT (\% of children ages 12-23 months)  & Yes & No & 0.81 & 0.23 \\ 
P04\_11 & Life expectancy at birth, total (years)  & No & No & 0.69 & 0.23 \\ 
P04\_12 & Survival to age 65, female (\% of cohort)  & No & No & 0.77 & 0.22 \\ 
P04\_13 & Survival to age 65, male (\% of cohort)  & No & No & 0.70 & 0.22 \\ 

\\
\multicolumn{2}{l}{\textbf{Education}}\\
P05\_1 & Quality of primary education & No & No & 0.43 & 0.20 \\ 
P05\_2 & Quality of math and science education & No & No & 0.50 & 0.19 \\ 
P05\_3 & Extent of staff training & No & No & 0.54 & 0.18 \\ 

\\
\multicolumn{2}{l}{\textbf{Goods market efficiency}}\\
P06\_1 & Intensity of local competition & No & No & 0.63 & 0.17 \\ 
P06\_2 & Extent of market dominance & No & No & 0.44 & 0.20 \\ 
P06\_3 & Effectiveness of anti-monopoly policy & No & No & 0.48 & 0.20 \\ 
P06\_5 & Agricultural policy costs & No & No & 0.50 & 0.13 \\ 
P06\_7 & Degree of customer orientation & No & No & 0.59 & 0.16 \\ 
P06\_8 & Buyer sophistication & No & No & 0.42 & 0.19 \\ 

\\
\multicolumn{2}{l}{\textbf{Labor market efficiency}}\\
P07\_1 & Cooperation in labor-employer relations & No & No & 0.51 & 0.17 \\ 
P07\_2 & Redundancy costs, weeks of salary*  & Yes & Yes & 0.65 & 0.29 \\ 
P07\_4 & Pay and productivity & No & No & 0.48 & 0.17 \\ 
P07\_5 & Reliance on professional management & No & No & 0.55 & 0.20 \\ 
P07\_8 & Labor force participation rate for ages 15-24, total (\%) (modeled ILO estimate)  & No & No & 0.48 & 0.21 \\ 

\\
\multicolumn{2}{l}{\textbf{Financial market development}}\\
P08\_3 & Financing through local equity market & No & No & 0.51 & 0.20 \\ 
P08\_4 & Ease of access to loans & No & No & 0.40 & 0.19 \\ 
P08\_5 & Venture capital availability & No & No & 0.35 & 0.19 \\ 
P08\_6 & Soundness of banks & No & No & 0.69 & 0.16 \\ 
P08\_7 & Regulation of securities exchanges & No & No & 0.58 & 0.18 \\ 

\\
\multicolumn{2}{l}{\textbf{Technological readiness}}\\
P09\_1 & Availability of latest technologies & No & No & 0.60 & 0.20 \\ 
P09\_2 & Firm-level technology absorption & No & No & 0.54 & 0.19 \\ 
P09\_3 & FDI and technology transfer & No & No & 0.57 & 0.16 \\ 

\\
\multicolumn{2}{l}{\textbf{Business sophistication}}\\
P10\_1 & Local supplier quantity & No & No & 0.56 & 0.15 \\ 
P10\_2 & Local supplier quality & No & No & 0.53 & 0.18 \\ 
P10\_3 & State of cluster development & No & No & 0.49 & 0.21 \\ 
P10\_4 & Nature of competitive advantage & No & No & 0.38 & 0.23 \\ 
P10\_5 & Value chain breadth & No & No & 0.44 & 0.20 \\ 
P10\_6 & Control of international distribution & No & No & 0.52 & 0.18 \\ 
P10\_7 & Production process sophistication & No & No & 0.47 & 0.21 \\ 
P10\_8 & Extent of marketing & No & No & 0.52 & 0.20 \\ 
P10\_9 & Willingness to delegate authority & No & No & 0.43 & 0.18 \\ 

\\
\multicolumn{2}{l}{\textbf{R+D Innovation}}\\
P11\_1 & Capacity for innovation & No & No & 0.42 & 0.21 \\ 
P11\_2 & Quality of scientific research institutions & No & No & 0.49 & 0.20 \\ 
P11\_3 & Company spending on R+D & No & No & 0.38 & 0.20 \\ 
P11\_4 & University-industry collaboration in R+D & No & No & 0.47 & 0.21 \\ 
P11\_5 & Government procurement of advanced tech. products & No & No & 0.43 & 0.14 \\ 
P11\_6 & Availability of scientists and engineers & No & No & 0.49 & 0.19 \\ 
P11\_7 & Intellectual property protection & No & No & 0.45 & 0.23 \\ 

\\
\multicolumn{2}{l}{\textbf{Public Governance}}\\
P12\_1 & Control of Corruption & No & No & 0.45 & 0.23 \\ 
P12\_2 & Government Effectiveness & No & No & 0.55 & 0.19 \\ 
P12\_3 & Regulatory Quality & No & No & 0.60 & 0.18 \\ 
P12\_4 & Rule of Law & No & No & 0.58 & 0.20 \\ 
P12\_5 & Voice and Accountability & No & No & 0.58 & 0.22 \\ 
P12\_6 & Property rights & No & No & 0.57 & 0.21 \\ 
P12\_7 & Diversion of public funds & No & No & 0.45 & 0.23 \\ 
P12\_8 & Public trust in politicians & No & No & 0.33 & 0.23 \\ 
P12\_9 & Judicial independence & No & No & 0.50 & 0.24 \\ 

\\
\multicolumn{2}{l}{\textbf{Cost of doing business}}\\
P13\_1 & Cost of business start-up procedures (\% of GNI per capita)  & Yes & Yes & 0.86 & 0.22 \\ 
P13\_2 & Time required to enforce a contract (days)  & No & Yes & 0.70 & 0.17 \\ 
P13\_3 & Time required to register property (days)  & Yes & Yes & 0.82 & 0.18 \\ 
P13\_4 & Time required to start a business (days)  & Yes & Yes & 0.77 & 0.21 \\ 
P13\_5 & Time to resolve insolvency (years)  & No & Yes & 0.75 & 0.14 \\ 
P13\_6 & Business costs of terrorism & No & No & 0.71 & 0.18 \\ 
P13\_7 & Business costs of crime and violence & No & No & 0.56 & 0.22 \\